\let\PoreTimeOriginalBibliographyStyle\bibliographystyle
\renewcommand{\bibliographystyle}[1]{\PoreTimeOriginalBibliographyStyle{sn-mathphys-num-articleids}}
\documentclass[pdflatex,sn-mathphys-num]{sn-jnl}
\let\bibliographystyle\PoreTimeOriginalBibliographyStyle

\usepackage{graphicx}%
\usepackage{multirow}%
\usepackage{amsmath,amssymb,amsfonts}%
\usepackage{amsthm}%
\usepackage{mathrsfs}%
\usepackage[title]{appendix}%
\usepackage{xcolor}%
\usepackage{textcomp}%
\usepackage{manyfoot}%
\usepackage{booktabs}%
\usepackage{algorithm}%
\usepackage{algorithmicx}%
\usepackage{algpseudocode}%
\usepackage{listings}%
\usepackage{subcaption}
\usepackage{placeins}

\usepackage{booktabs}

\theoremstyle{thmstyleone}%
\theoremstyle{thmstyletwo}%

\theoremstyle{thmstylethree}%

\unnumbered

\usepackage{array,tabularx,ragged2e,float,needspace}
\newcolumntype{L}[1]{>{\raggedright\arraybackslash\hspace{0pt}}p{#1}}
\newcolumntype{Y}{>{\raggedright\arraybackslash\hspace{0pt}}X}
\newcounter{suppnote}
\newcommand{\suppnote}[1]{%
  \Needspace{5\baselineskip}%
  \refstepcounter{suppnote}%
  \subsection*{Supplementary Note \thesuppnote. #1}%
  \addcontentsline{toc}{subsection}{Supplementary Note \thesuppnote. #1}%
}
\hypersetup{pdftitle={Spatiotemporal learning improves quantification of pore-space evolution and fluid displacement},pdfauthor={Zhuangzhuang Ma, Rukuan Chai, Qianqian Ma, Yanghua Wang, Branko Bijeljic, Martin J. Blunt}}
\begin{document}

\title[Pore-space evolution and fluid displacement]{Spatiotemporal learning improves quantification of pore-space evolution and fluid displacement}

\author[1,2]{\fnm{Zhuangzhuang} \sur{Ma}}

\author*[2]{\fnm{Rukuan} \sur{Chai}}\email{r.chai@imperial.ac.uk}

\author[1,2]{\fnm{Qianqian} \sur{Ma}}

\author[1,2]{\fnm{Yanghua} \sur{Wang}}

\author[2]{\fnm{Branko} \sur{Bijeljic}}

\author*[2]{\fnm{Martin J.} \sur{Blunt}}\email{m.blunt@imperial.ac.uk}

\affil[1]{\orgdiv{Resource Geophysics Academy}, \orgname{Imperial College London}, \orgaddress{\city{London}, \postcode{SW7 2BP} \country{United Kingdom}}}

\affil*[2]{\orgdiv{Department of Earth Science and Engineering}, \orgname{Imperial College London}, \orgaddress{\city{London}, \postcode{SW7 2AZ}, \country{United Kingdom}}}

\abstract{
Pore-space evolution and fluid displacement influence how rocks transmit and store fluids, affecting geological carbon storage, underground hydrogen storage and groundwater contaminant transport. A single time-resolved micro-CT experiment may produce tens to hundreds of three-dimensional images, and segmentation is required to quantify physical properties and run image-based simulations. Traditional methods such as thresholding perform poorly on noisy, low-contrast images and need manual adjustment per image; 3D deep learning networks segment each image without temporal information, so their errors vary between images and could be mistaken for physical change. Here we introduce a transformer-based 4D segmentation framework combining spatial and temporal information. It uses sequential images to resolve ambiguous phase assignments and separate physical change from segmentation error. The method is tested on images of mineral dissolution: it measures pore-volume growth and rock boundaries more accurately than an equivalent 3D model that ignores temporal information. It also captures the increase in permeability verified against reference segmentations and flow simulations. The advantage transfers to another dissolution sample without retraining. On two other datasets of fluid displacement, it reduces spurious changes in predicted rock and porosity fluctuations and accurately identifies filling events. Spatiotemporal segmentation turns sequences of hundreds of three-dimensional images into more reliable quantitative measurements of pore-space evolution and fluid displacement without per-image parameter adjustment to support the analysis of time-resolved CT imaging.
}

\maketitle

\section{Introduction}\label{sec1}

Fluid transport and reactions in porous rocks are central to geological carbon dioxide storage, underground hydrogen storage and contaminant transport in groundwater~\cite{bachu2000sequestration,chai2022co2,ZIVAR202123436,honeyman1999colloidal}. In geological carbon storage, capillary forces retain carbon dioxide in pores and dissolution transfers it into formation water~\cite{mingotti2025convective}. Fluid filling and snap-off, in which a fluid connection breaks, can reorganise the pathways available for flow~\cite{berg2013real,singh2017dynamics}. The solid structure can also evolve: mineral dissolution can change pore geometry and rock permeability~\cite{menke2015dynamic,ma2026time,salala2025biobased}. Quantifying these processes requires accurate measurements of how fluid occupancy and pore geometry evolve over time.

Time-resolved X-ray micro-tomography provides three-dimensional (3D) images of these evolving rock--fluid systems~\cite{berg2013real,andrew2015imaging,bultreys2015real,menke2015dynamic,singh2017dynamics}. Each 3D image corresponds to a time step in the image sequence. Segmentations of successive images provide porosity, the fraction of a volume occupied by pores, and saturation, the fraction of pore space occupied by a given fluid. Segmented images enable measurements of pore structure and fluid occupancy and provide computational domains for flow and reactive transport simulations~\cite{maes2021geochemfoam}. The accuracy of phase volumes and boundaries therefore matters both for measuring changes and for the physical calculations based on them~\cite{gao2024gradient}.

Segmentation assigns each voxel, a 3D image element, to a material phase. Errors that vary between images can distort measurements of changes in pore space over time and obscure fluid displacement. Greyscale values can also vary between successive images even when the material remains unchanged, complicating consistent phase identification across the sequence~\cite{zhang2019challenges}. Fluid motion during a CT scan can cause temporal blurring: projections acquired at different stages of displacement are combined in one reconstruction, producing intermediate greyscale values~\cite{goethals2022dynamic,gao2017intermittency}. When these intermediate greyscale values resemble those of rock, segmentation can misclassify a fluid-filled pore as rock. If the same pore is correctly identified in a later image, the segmentation suggests that new pore space has formed, although only the fluid occupying the pore has changed. In laboratory CT, shorter duration scans can increase image noise, making phase identification more difficult as temporal resolution improves~\cite{vanoffenwert2019pore,goethals2022dynamic,tekseth2024imaging}. Noise and weak phase contrast complicate thresholding and watershed segmentation, while manual adjustment for each image becomes laborious across sequences of tens to hundreds of images~\cite{iassonov2009segmentation,schluter2014image}. For example, Tang et al. reported that a conventional workflow involving registration, segmentation and morphological adjustment required approximately 30 hours to process 20 3D images~\cite{tang2025scaling}. Annotation-free supervoxel methods offer rapid segmentation of noisy multiphase images~\cite{ye2025noise}. Deep learning has improved the spatial resolution of multiphase images~\cite{ma2025super} and their segmentation~\cite{niu2020digital,da2021deep_Unet,zhang2026samamba3d}. However, 3D networks that process each image independently do not use information from other images in the time sequence. Using this information offers a way to resolve ambiguous phase assignments across the sequence. The benefit of temporal information should therefore be assessed through both segmentation accuracy and the resulting measurements of pore geometry and fluid occupancy.

Here we introduce PoreTimeNet, a spatiotemporal (4D) segmentation network that uses successive images to improve measurements of pore-space evolution and fluid displacement. We compare it with a 3D network that uses the same spatial architecture but processes each image independently. During carbonate dissolution, the 4D model identified rock boundaries more accurately and reduced mean absolute porosity error from 0.97 to 0.28 percentage points. The resulting segmentations also gave more accurate estimates of the increase in permeability. During drainage and imbibition, the 4D model reduced spurious changes in predicted rock between consecutive images by approximately one-third compared with the 3D model. In low-contrast laboratory CT images, the 4D model identified pores with intermediate greyscale values that the 3D model misclassified as rock, allowing more of the observed oil filling to be measured. 

\section{Results}
\label{sec3}

\subsection{A framework for time-resolved image segmentation}
\label{sec:framework_overview}

\begin{figure}[!htbp]
\centering
\includegraphics[width=\textwidth,height=0.95\textheight,keepaspectratio]{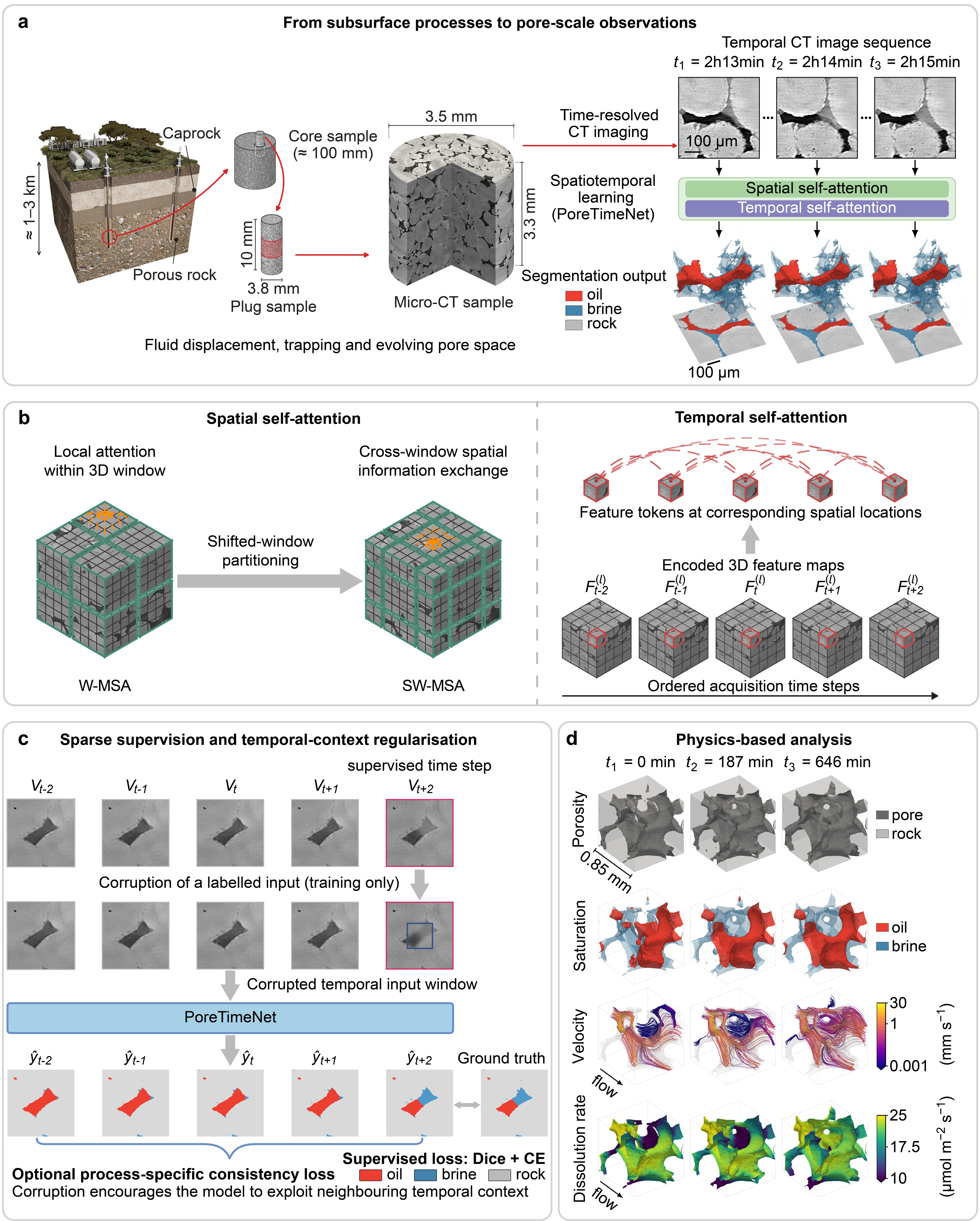}
\caption{\textbf{A framework for measuring evolving pore geometry and fluid occupancy.}
\textbf{a}, Subsurface processes motivate time-resolved imaging of rock samples. 
\textbf{b}, Spatial attention combines spatial information within each image; temporal attention combines information from other images in the time sequence.
\textbf{c}, Training with labels at selected time steps. Input corruption and a process-specific consistency loss are optional settings.
\textbf{d}, Segmented geometry provides porosity and saturation measurements and domains for flow and reactive-transport simulations.}
\label{fig:framework}
\end{figure}

The framework takes an aligned sequence of three-dimensional greyscale images and assigns each voxel to specified phases (Fig.~\ref{fig:framework}). A shared spatial encoder describes local geometry, while temporal attention and decoder fusion combine information from other images in the time sequence. The spatial encoder and decoder follow the SwinUNETR design~\cite{hatamizadeh2021swin}; the 3D SwinUNETR model is referred to as the 3D model throughout. During training, reference segmentations supervise predictions at the labelled time steps within each input window. Other images in the same input window provide temporal information even without training labels. Overlapping spatial predictions are combined to cover the evaluation volume. 

Evaluation connects phase and boundary accuracy, consistency through time, and physical calculations from the segmented geometry. Together, these measurements assess both the predictions and their consequences for inferred rock properties.

\subsection{Measuring pore-space evolution during dissolution}
\label{sec:dissolution_measurements}

\begin{figure}[!htbp]
\centering
\includegraphics[width=\textwidth,height=0.67\textheight,keepaspectratio]{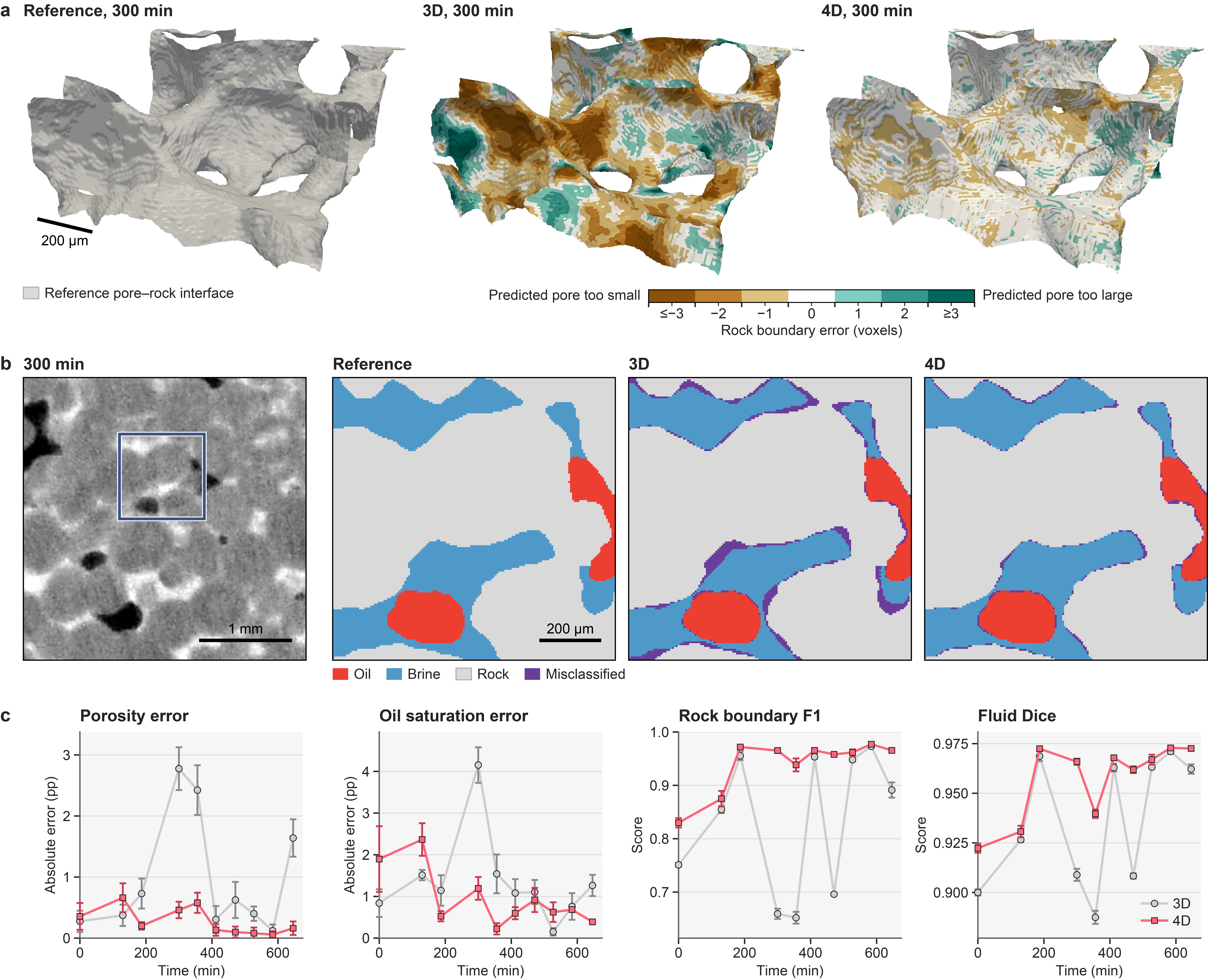}
\caption{\textbf{Measuring pore-space evolution during carbonate dissolution.}
\textbf{a}, Pore--rock surfaces at 300\,min. Predicted surfaces are coloured by signed distance to the reference boundary: brown indicates locally smaller pores and green larger pores.
\textbf{b}, A greyscale cross-section and segmentations of the boxed region. Purple marks disagreement between the predictions and the reference segmentation, including misclassification between oil and brine.
\textbf{c}, Absolute porosity and oil-saturation errors, rock boundary F1 score, and fluid Dice score across all ten images with reference segmentations. At each time step, points show means and error bars show $\pm1$ sample standard deviation across three training runs with different random seeds. Error bars smaller than their markers are not visible.}
\label{fig:dissolution_measurements}
\end{figure}

We tested whether the 4D model could measure pore-volume growth and identify rock boundaries more accurately than the 3D model. The experiment followed oil displacement by CO$_2$-saturated brine as the carbonate dissolved~\cite{ma2026time}. Predictions were compared with reference segmentations at ten time steps (Fig.~\ref{fig:dissolution_measurements}; Methods).

The 4D model tracked pore-volume evolution more closely than the 3D model. Mean absolute porosity error decreased from 0.97 to 0.28 percentage points. The error in predicted oil-saturation also decreased. 

Better pore-volume measurement accompanied more accurate rock boundaries. The mean rock boundary F1 (BF1) score increased from 0.833 to 0.941. Segmentation scores are given in Supplementary Table~\ref{supp-tab:dissolution_temporal_ablation}. At brine--rock boundaries, where the two phases have similar greyscale values and are difficult to distinguish, PoreTimeNet learned features that distinguished the two phases more clearly than the 3D model did (Supplementary Fig.~\ref{supp-fig:decoder_features} and Supplementary Note~\ref{supp-sec:decoder_features}).

The surface and slice views show how these errors affect the reconstructed pore boundary (Fig.~\ref{fig:dissolution_measurements}a,b; Supplementary Note~\ref{supp-sec:supp_temporal_ablation_results}). As an independent baseline, we used nnU-Net~\cite{isensee2021nnu}, a self-configuring 3D network trained with its own configuration and training procedure on the same training region and ten labelled time steps. Compared with 4D, nnU-Net had larger rock-boundary, porosity and oil-saturation errors (Supplementary Note~\ref{supp-sec:nnunet_ct} and Supplementary Table~\ref{supp-tab:nnunet_ct_summary}), as well as a larger error in the estimated permeability increase.

\subsection{From segmented geometry to flow and transport}
\label{sec:segmentation_derived_properties}

\begin{figure}[!htbp]
\centering
\includegraphics[width=\textwidth]{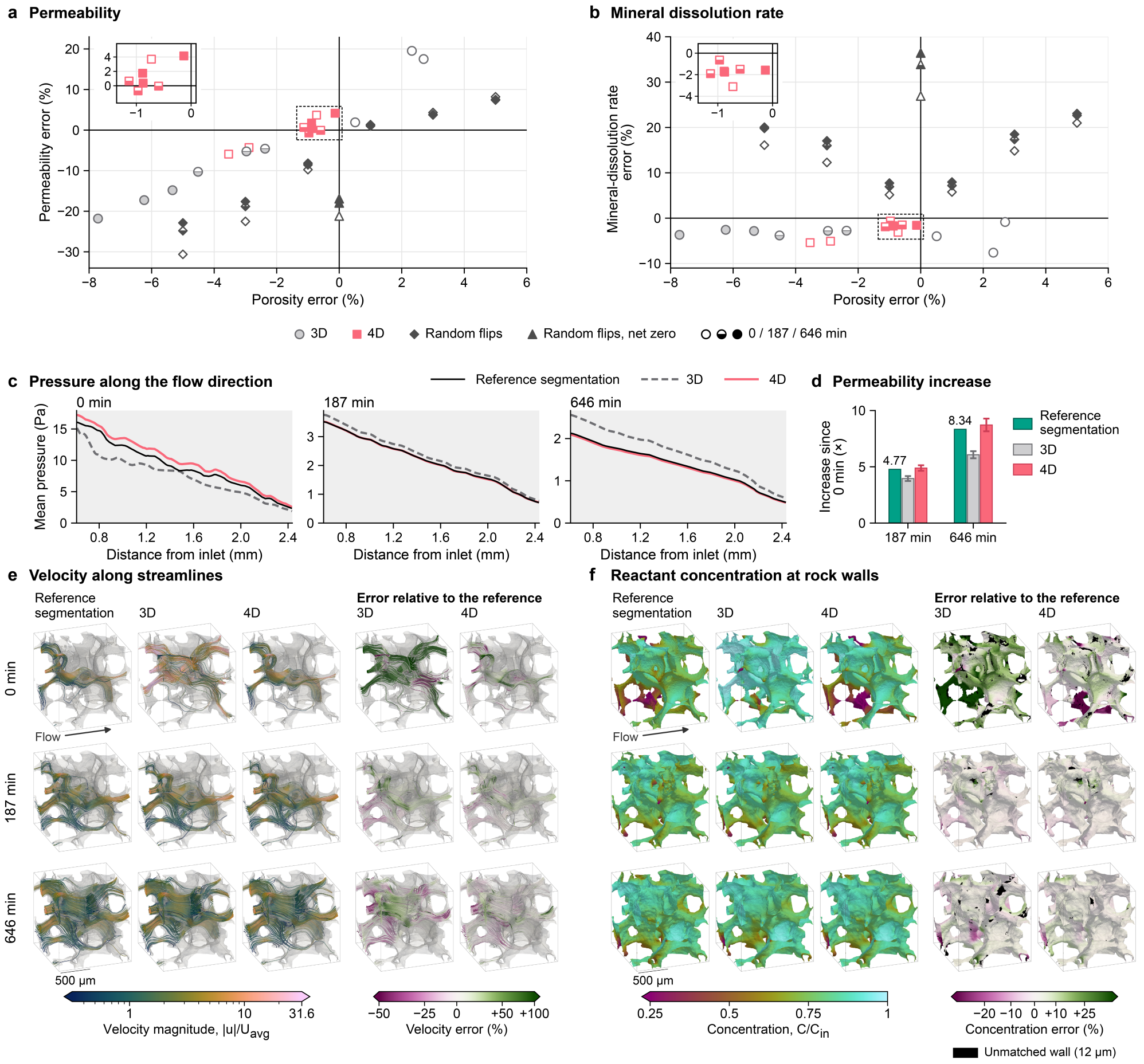}
\caption{\textbf{Effects of segmentation on simulated flow and mineral dissolution rate.}
\textbf{a,b}, Relative errors in permeability and total mineral dissolution rate versus relative porosity error, calculated as $100(X/X_{\mathrm{ref}}-1)$, where $X_{\mathrm{ref}}$ is obtained from the reference segmentation at the same time step. Each model point represents one training run. Random-perturbation points show means, with ranges in Supplementary Table~\ref{supp-tab:random_boundary_ranges}. Net-zero perturbations preserve porosity by converting equal numbers of rock voxels to pore and pore voxels to rock.
\textbf{c}, Pressure averaged over pore voxels in each cross-section along the flow direction. The plotted interval (0.61--2.44\,mm from the inlet) was used to calculate permeability. Pressure was set to zero at the outlet of the full simulation domain.
\textbf{d}, Permeability divided by its value at 0\,min, $K(t)/K(0)$. Model bars show means and error bars show $\pm1$ sample standard deviation across three training runs with different random seeds.
\textbf{e,f}, Velocity magnitude along streamlines and concentration at rock surfaces. Velocity is normalised by the mean pore velocity in each simulation; concentration by the inlet concentration. Velocity errors compare values at the same spatial position; concentration errors compare values at the nearest corresponding rock-surface points. Black surface regions have no reference surface within approximately 12\,$\mu$m and are excluded from the concentration-error comparison.}
\label{fig:flow_simulation}
\end{figure}

We examined how differences in segmented pore geometry affect the calculated permeability and total mineral dissolution rate. We used single-phase flow and reactive-transport simulations for each segmentation at 0, 187 and 646\,min, corresponding to the initial state and the ends of the two rapid-dissolution stages~\cite{ma2026time} (Fig.~\ref{fig:flow_simulation}). The pore space comprised all voxels assigned to oil or brine and was assumed to be fully saturated with a single fluid. All simulations used the original voxel grid and the same physical parameters. Their solutions passed common checks of flow convergence, mass balance and concentration stability (Methods).

The 3D model underestimated the final-to-initial permeability ratio, $K(646)/K(0)$, by 27\%. The 4D model estimated this ratio with a mean absolute relative error of 6.7\% (Fig.~\ref{fig:flow_simulation}d). The 3D model overestimated permeability at 0\,min by 13\% and underestimated it at 646\,min by 18\%, both contributing to the lower ratio. nnU-Net underestimated the ratio by 41\%, with permeability overestimated by 31\% at 0\,min and underestimated by 23\% at 646\,min (Supplementary Table~\ref{supp-tab:nnunet_ct_summary}).

The 4D model estimated pore-volume growth more accurately than the 3D model and nnU-Net. At 646\,min, pore volume was 1.71 times its initial value in the reference segmentation, compared with 1.74 for 4D, 1.57 for 3D and 1.53 for nnU-Net.

Across the three time steps (0, 187 and 646 min), the mean absolute relative error in mineral dissolution rate was 3.4\% for the 3D model and 2.5\% for the 4D model (Fig.~\ref{fig:flow_simulation}b). Values at each time step are reported in Supplementary Table~\ref{supp-tab:cfd_model_values}.

We tested two types of random perturbation at the reference pore boundaries: one changed porosity, and the other preserved it. In the first, we converted rock voxels to pore or pore voxels to rock, increasing or decreasing pore volume by 1, 3 or 5\%. In the second, we converted 3\% of the original pore voxels to rock and the same number of rock voxels to pore, leaving porosity unchanged. Despite the unchanged porosity, permeability decreased by 17--21\% and the mineral dissolution rate increased by 27--36\% across the three time steps (Fig.~\ref{fig:flow_simulation}a,b; Supplementary Methods). These perturbations show that permeability and mineral dissolution rate depend on pore-boundary geometry as well as pore volume, and that the same boundary changes can affect the two properties in opposite directions. A segmented image used for flow and reactive-transport simulations should therefore be evaluated by the permeability and dissolution rate calculated from it, not by its phase volumes alone.

\subsection{Contribution of information from other images}
\label{sec:temporal_information_controls}

\begin{figure}[!htbp]
\centering
\includegraphics[width=\textwidth,height=0.67\textheight,keepaspectratio]{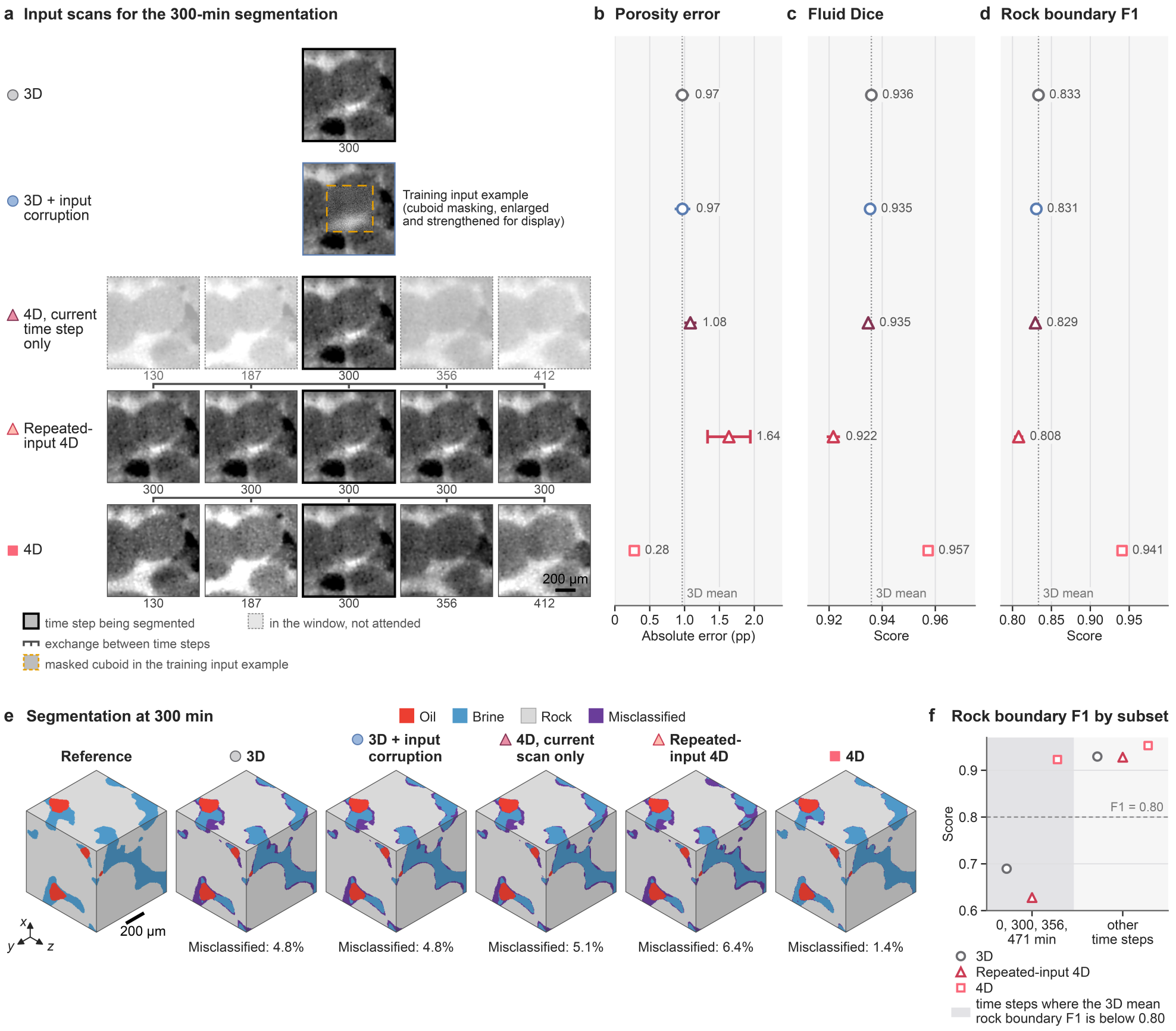}
\caption{\textbf{The contribution of other images in the time sequence to dissolution segmentation.}
\textbf{a}, Different model inputs for the prediction at 300\,min. The 3D model was retrained with input corruption; the masking example is enlarged for display. The current-time-step-only 4D model processed each image independently during training and inference. Repeated-input 4D uses the trained 4D model with repeated copies of the image at the current time step during inference.
\textbf{b--d}, Mean absolute porosity error, fluid Dice and rock boundary F1 scores, averaged over all ten time steps within each training run. Points show means and error bars show $\pm1$ sample standard deviation across three training runs with different random seeds. Dotted lines mark the mean for the 3D model. Error bars smaller than their markers are not visible.
\textbf{e}, Segmentation faces of the same $128^3$-voxel region at 300\,min. Purple marks misclassified voxels.
\textbf{f}, Rock boundary F1 scores for the four time steps where the 3D mean is below 0.80 and for the remaining six.}
\label{fig:input_modes}
\end{figure}

We performed three comparisons to test whether the 4D model's performance gain was from the incorporation of spatial and temporal information from other images in the time sequence (Fig.~\ref{fig:input_modes}). PoreTimeNet achieved its largest gains in boundary accuracy in the four images where the 3D model performed poorly, increasing mean rock BF1 score from 0.690 to 0.923 (Fig.~\ref{fig:input_modes}f).

First, we tested whether input corruption, an additional form of training data augmentation, could produce similar improvements in the 3D model. During training, we added noise, masked image regions or randomly varied the greyscale values of a selected training image. The corresponding reference segmentation remained unchanged. Porosity error, fluid Dice and rock BF1 scores remained close to those of the original 3D model.

Second, we trained a current-time-step-only 4D model that processed each image independently during training and inference. It retained the architecture and nominal parameter count of the full 4D model, but its accuracy remained close to that of the 3D model (Supplementary Table~\ref{supp-tab:dissolution_temporal_ablation}).

Third, during inference, we repeated the current image multiple times to form the input sequence, keeping the trained 4D model's weights unchanged. Mean rock BF1 scores fell from 0.941 to 0.808, and the porosity error increased. The largest performance drops occurred at the time steps where the 3D model had the lowest boundary accuracy (Fig.~\ref{fig:input_modes}f).

Together, these comparisons show that the 4D model estimates porosity and identifies phase boundaries more accurately because it uses information from other images in the time sequence through temporal attention and fusion, rather than because of input corruption or a larger network alone.

The 4D model also reduced pore-to-rock changes in its predictions during dissolution. Predicting rock in a region previously identified as pore space can be mistaken for precipitation. Temporal majority voting across three consecutive time steps further reduced these changes but made fluid identification less accurate, whereas the 4D model improved both consistency and accuracy. These results show that fewer pore-to-rock changes in the predictions do not necessarily indicate a more accurate representation of pore-space evolution: temporal consistency needs to be assessed together with phase-identification accuracy (Supplementary Note~\ref{supp-sec:d2_consistency} and Supplementary Table~\ref{supp-tab:d2_consistency}).

\subsection{Rock-geometry consistency during fluid displacement}
\label{sec:ketton_matched_displacement}

We then studied another dataset on Ketton limestone, where synchrotron X-ray imaging was used to observe drainage and imbibition without any dissolution~\cite{singh2018time}. The 4D model maintained the rock geometry more consistently as fluids moved through it. Training and evaluation settings are described in Supplementary Note~\ref{supp-sec:ketton_matched}.

\begin{figure}[!htbp]
\centering
\includegraphics[width=\textwidth,height=0.62\textheight,keepaspectratio]{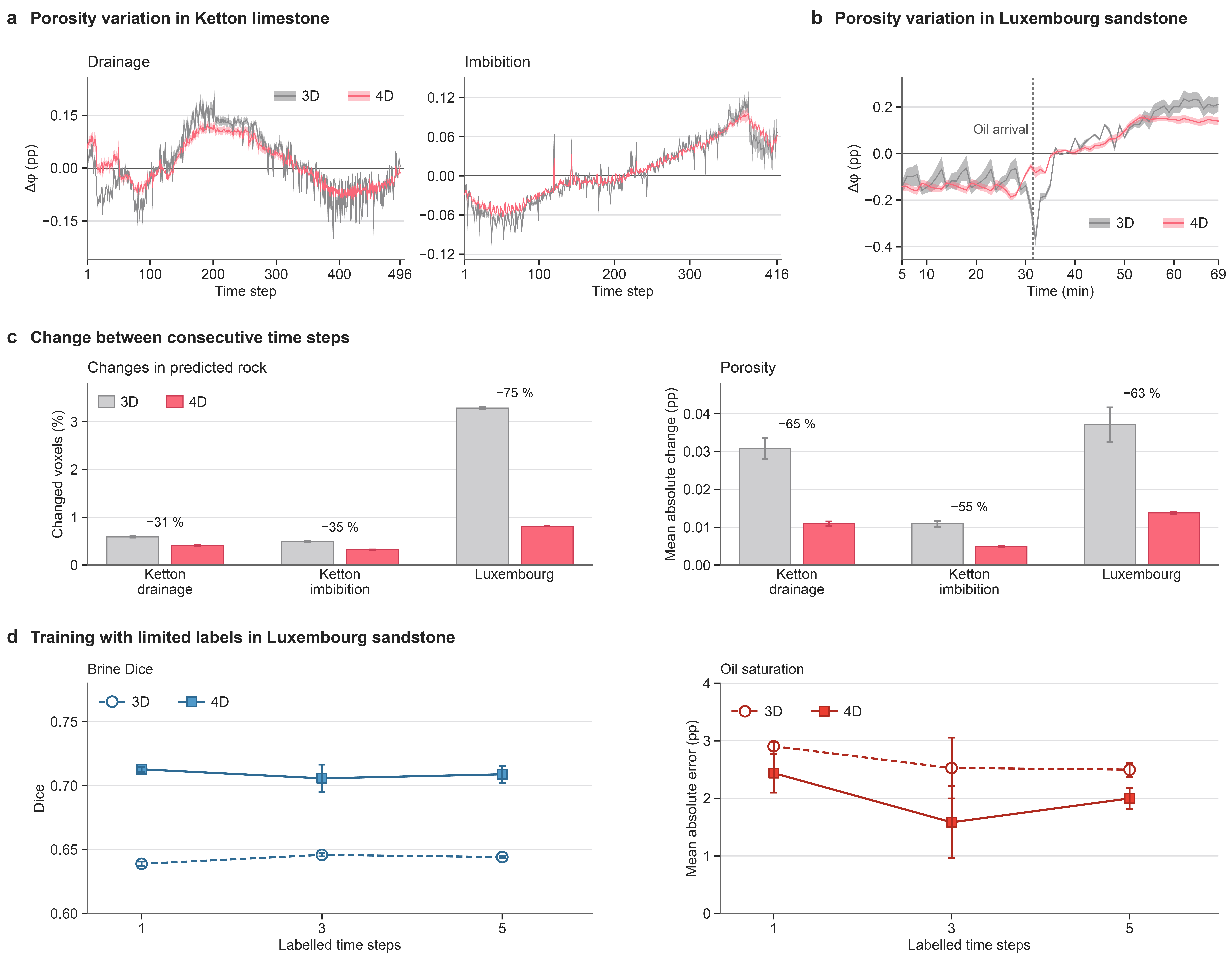}
\caption{\textbf{Temporal consistency of the solid geometry and segmentation with limited training labels.}
\textbf{a,b,} Porosity deviation, $\Delta\phi$, in percentage points (pp), for Ketton drainage and imbibition (a) and Luxembourg sandstone (b). Medians use the complete Ketton sequences and time steps at 5--69\,min for Luxembourg sandstone. The temporal median of each run was subtracted before averaging across training runs. The sandstone interval excludes sequence ends, where input windows repeat the image at the nearest endpoint. The dotted line in b marks oil arrival in the evaluation region. The sandstone panels b and c use models trained with one labelled training time step.
\textbf{c,} Changes in predicted rock voxels between consecutive time steps (left), measured as the percentage of evaluated voxels whose predictions changed between rock and pore, and mean absolute porosity change (right). Statistics use pairs ending at time steps 26--471 for drainage and 26--391 for imbibition. The later time step in each sandstone pair is at 5--69\,min, including the 4--5-min pair. Printed percentages give the relative change of the 4D mean from the 3D mean.
\textbf{d,} Luxembourg sandstone brine Dice score and mean absolute oil-saturation error with one, three or five labelled training time steps, averaged over the designated evaluation times, excluding the 37-min time step. Blue and red denote brine and oil; hollow circles with dashed lines denote 3D, and filled squares with solid lines denote 4D. Curves and bars show means over three training seeds; shading and error bars indicate $\pm1$ sample standard deviation.}
\label{fig:temporal_consistency}
\end{figure}

Changes in predicted rock voxels between consecutive time steps decreased from 0.590\% in 3D to only 0.409\% for 4D during drainage. They also decreased during imbibition, as did changes in porosity between time steps in both processes (Fig.~\ref{fig:temporal_consistency}a,c). These results show improved consistency in the segmented geometry of the fixed rock matrix.

Accuracy at time steps with reference segmentations improved more modestly. Fluid Dice and rock boundary F1 scores increased in both processes. Adding input corruption to 3D recovered much of the drainage accuracy gain, while 4D porosity predictions remained more consistent (Supplementary Tables~\ref{supp-tab:ketton_drainage} and~\ref{supp-tab:ketton_imbibition}).

Information from other images in the time sequence affected consistency more than the fluid Dice score against the reference segmentations. Replacing the trained 4D model's inputs with copies of the image at the current time step left the Dice score close to its original values but led to more changes in predicted rock at time steps 221--278 for drainage and 181--238 for imbibition (Supplementary Table~\ref{supp-tab:ketton_repeated}).

We also compared changes between predicted oil and brine volumes. Adding the volumes that changed from brine to oil and from oil to brine gave the total change; subtracting the latter from the former gave the net change. In both processes, the total was lower for 4D, while the net change remained close to the 3D estimates (Fig.~\ref{fig:pore_dynamics}b).

The final test was a fast laboratory CT displacement sequence in Luxembourg sandstone~\cite{mascini2021fluid} in which brine and rock were difficult to distinguish; models were trained with annotations from one time step. The 4D predictions also maintained the fixed rock geometry more consistently. For pairs of time steps with the later one at 5--69\,min, changes in predicted rock decreased from 3.28\% of valid voxels for 3D to 0.81\% for 4D. Changes in porosity between time steps were also smaller (Fig.~\ref{fig:temporal_consistency}b,c).

\subsection{Measuring changes in fluid occupancy}
\label{sec:luxembourg_limited_labels}

Shorter scans can increase image noise, making reliable phase identification a challenge for faster laboratory CT imaging~\cite{goethals2022dynamic}. We used the Luxembourg sandstone sequence to examine whether temporal information could improve measurements of the locations and volumes of oil filling.

\begin{figure}[!htbp]
\centering
\includegraphics[width=\textwidth,height=0.60\textheight,keepaspectratio]{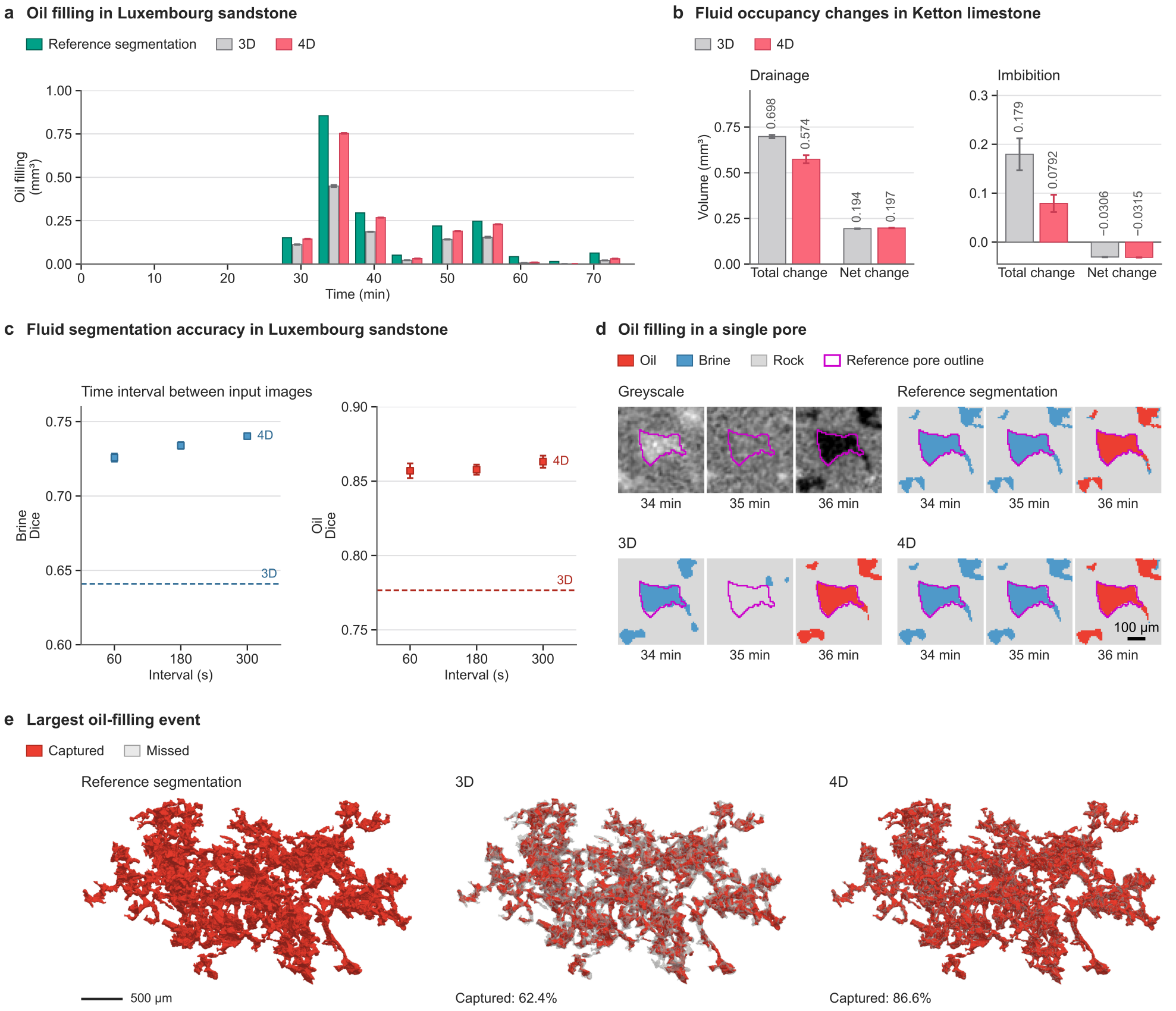}
\caption{\textbf{Fluid occupancy changes and their recovery from dynamic images.}
\textbf{a,} Luxembourg sandstone oil-filling volumes: connected brine-to-oil changes over sampled 5- or 6-min intervals. Time steps give scan start times relative to continuous imaging.
\textbf{b,} Ketton limestone total and net change: the sum and difference, respectively, of brine-to-oil and oil-to-brine transition volumes over consecutive time steps. Volumes were summed over consecutive image pairs within each training run and then averaged across the three runs.
\textbf{c,} Brine and oil Dice scores at common evaluation time steps when the input images were separated by 60, 180 or 300\,s. Squares denote 4D; dashed lines denote 3D.
\textbf{d,} During the scan at 35\,min, a filling pore in the sandstone resembles rock because of intermediate greyscale values. The 3D prediction assigns it to rock; the reference segmentation and 4D prediction assign it to brine. Magenta outlines the reference pore.
\textbf{e,} The largest reference filling region, at 32--37\,min. Red and grey show its volume captured and missed by predicted brine-to-oil changes. Renderings show one seed; percentages average three. Means and shading or error bars represent three seeds and $\pm1$ sample standard deviation. Filtering and recovery calculations are described in Supplementary Note~\ref{supp-sec:lux_observed_filling}.}
\label{fig:pore_dynamics}
\end{figure}

The 4D model estimated oil-filling volume more accurately. We measured oil filling as the volume of pore space that changed from brine to oil over each selected time interval (Fig.~\ref{fig:pore_dynamics}a). The total oil-filling volume across these intervals was 1.94\,mm$^3$ in the reference segmentations, compared with 1.09\,mm$^3$ for the 3D predictions and 1.66\,mm$^3$ for the 4D predictions (Fig.~\ref{fig:pore_dynamics}a; Supplementary Table~\ref{supp-tab:lux_filling_volume}). Most filling missed by 3D (about 72\%) occurred where the earlier prediction was rock and the later prediction was oil. Such a change is not counted as brine-to-oil filling.

A local pore illustrates how this error arose (Fig.~\ref{fig:pore_dynamics}d). During the scan for the 35-min time step, the filling pore resembled rock in the greyscale image. Fluid displacement during a CT scan can produce intermediate greyscale values because the reconstruction combines projections acquired at different stages of the displacement event~\cite{goethals2022dynamic}. The 3D models classified most of the pore as rock, whereas the 4D models identified it as pore space. Both recognised oil in the following image. The earlier misclassification therefore caused the 3D segmentation to miss filling that the 4D model captured. The 4D predictions also captured more of the largest connected reference filling region (Fig.~\ref{fig:pore_dynamics}e).

We then examined how often this misclassification occurred across individual pores. The reference segmentations identified 1,603 pores of at least 100 voxels that showed brine-to-oil changes between adjacent time steps during 5--69\,min. In the image preceding the brine-to-oil change, many of these pores closely resembled rock, with greyscale values similar to or lower than those of rock. These pores accounted for 54\% of the total volume of the 1,603 pores analysed (Supplementary Fig.~\ref{supp-fig:lux_midscan_filling} and Note~\ref{supp-sec:lux_midscan_filling}).

The 3D model classified each image independently, without information from other images in the time sequence, and performed poorly in distinguishing these pores from rock. It misclassified 90\% of the volume of these rock-like pores as rock in the preceding image, approximately 4.7 times the fraction misclassified by the 4D model. Simple temporal voting could not resolve this misclassification. For one 3D training run, we assigned each voxel the majority prediction across five consecutive images, centred on the image preceding the brine-to-oil change. After voting, 91\% of the volume in the rock-like pores was still incorrectly classified as rock (Supplementary Table~\ref{supp-tab:lux_midscan_rock}).

When both models were trained with one labelled time step, the 4D model achieved a brine Dice score of 0.713, compared with 0.639 for the 3D model (Fig.~\ref{fig:temporal_consistency}d). The 4D model also had lower oil-saturation error (Supplementary Tables~\ref{supp-tab:lux_budget} and~\ref{supp-tab:lux_budget_volume}).

The 4D model retained higher brine and oil Dice scores than the 3D model when the input images were separated by 60, 180 or 300\,s (Fig.~\ref{fig:pore_dynamics}c; Supplementary Note~\ref{supp-sec:lux_input_spacing} and Supplementary Table~\ref{supp-tab:lux_input_spacing}). These results suggest potential for applying spatiotemporal segmentation to faster laboratory CT imaging, helping identify fluid phases and quantify their displacement at higher temporal resolution.

\subsection{Transfer to other dissolution sequences}
\label{sec:dissolution_transfer}

We applied the models trained on the dissolution sequence described above (Seq1) without fine-tuning or target-based checkpoint selection to mixed-wet Seq2 and water-wet Seq3 from a second Ketton limestone sample~\cite{ma2026multiphase}. This comparison assessed segmentation agreement using phase Dice and boundary F1 scores (Table~\ref{tab:dissolution_transfer}; Supplementary Note~\ref{supp-sec:transfer_matched}).

The 4D models retained a higher Dice score for each phase and higher boundary F1 scores for most phases in the new sequences. Oil boundary F1 score in Seq2 was slightly higher for 3D. The source-trained models therefore retained a temporal advantage in identifying rock and fluid regions across related dissolution experiments.

\begin{table}[htbp]
\centering\small
\setlength{\tabcolsep}{3pt}
\caption{\textbf{Segmentation transfer without fine-tuning.} Seq1-trained 3D and 4D models were applied to different experiments Seq2 and Seq3 without target adaptation or target-based checkpoint selection. Scores cover the full spatial extent of the reference segmentations within valid sample support. Values are means $\pm$ sample standard deviations across three seeds, after averaging the two time steps with reference segmentations within each seed. Fluid scores equally weight oil and brine. Unlike the central time-step output used for Seq1, 4D transfer predictions were combined across overlapping temporal windows using Gaussian weights.}
\label{tab:dissolution_transfer}
\begin{tabular}{lcccc}
\toprule
 & \multicolumn{2}{c}{Seq2 (mixed-wet)} & \multicolumn{2}{c}{Seq3 (water-wet)} \\
\cmidrule(lr){2-3}\cmidrule(lr){4-5}
Metric & 3D & 4D & 3D & 4D \\
\midrule
Rock Dice score & \(0.9672\pm0.0006\) & \(0.9763\pm0.0007\) & \(0.9513\pm0.0024\) & \(0.9555\pm0.0005\) \\
Oil Dice score & \(0.8258\pm0.0045\) & \(0.8275\pm0.0029\) & \(0.8516\pm0.0023\) & \(0.8678\pm0.0029\) \\
Brine Dice score & \(0.8709\pm0.0036\) & \(0.9042\pm0.0036\) & \(0.7552\pm0.0070\) & \(0.7745\pm0.0011\) \\
Fluid Dice score & \(0.8484\pm0.0007\) & \(0.8658\pm0.0027\) & \(0.8034\pm0.0030\) & \(0.8212\pm0.0017\) \\
Rock BF1 score & \(0.4637\pm0.0103\) & \(0.6087\pm0.0038\) & \(0.3661\pm0.0151\) & \(0.3803\pm0.0009\) \\
Oil BF1 score & \(0.5534\pm0.0173\) & \(0.5471\pm0.0036\) & \(0.5358\pm0.0018\) & \(0.5997\pm0.0059\) \\
Brine BF1 score & \(0.4567\pm0.0180\) & \(0.6201\pm0.0068\) & \(0.2753\pm0.0126\) & \(0.2952\pm0.0016\) \\
Fluid BF1 score & \(0.5050\pm0.0051\) & \(0.5836\pm0.0017\) & \(0.4056\pm0.0054\) & \(0.4474\pm0.0038\) \\
\bottomrule
\end{tabular}
\end{table}

\FloatBarrier
\section{Discussion}
\label{sec:discussion}

PoreTimeNet uses temporal information to improve measurements of pore-space evolution and fluid displacement. During dissolution, pore-volume growth quantifies the dissolution of solid material, while changes in pore-boundary geometry affect fluid flow. The 3D and 4D models were trained using the same annotations and training settings. The 3D model underestimated the final-to-initial permeability ratio, $K(646)/K(0)$, by approximately 27\%, whereas the 4D estimate was closer to the reference value. More accurate estimates of pore-volume growth and permeability changes help characterise how dissolution alters the relationship between porosity and permeability. The random boundary perturbations illustrated why segmentation evaluation needs both volume and boundary measurements. Exchanging equal numbers of rock and pore voxels preserved porosity while reducing permeability and increasing mineral dissolution rate. Differences in pore-boundary geometry can therefore affect both flow through the pore space and transport to reactive mineral surfaces. These perturbations complement phase-volume errors by showing how pore-boundary geometry affects calculated permeability and mineral dissolution rate.

The three comparisons during dissolution show that the 4D model's gains in porosity and boundary accuracy arise from information shared across distinct images. We also compared with another 3D method nnU-Net that again performed worse than PoreTimeNet.

In Luxembourg sandstone, improved brine identification recovered more of the observed oil filling and more of the largest connected region occupied by advancing oil. Reducing the misclassification of pores as rock in the earlier images preserves the brine-to-oil changes used to measure filling. The Ketton limestone displacement experiments complement this result by showing more consistent rock geometry during drainage and imbibition. The optional process-specific consistency loss shown in Fig.~\ref{fig:framework}c could incorporate constraints such as constant porosity during fluid displacement through a fixed rock matrix.

Spatiotemporal segmentation could support faster laboratory CT imaging by improving phase identification when individual images are ambiguous. In Luxembourg sandstone, other images in the time sequence helped distinguish pores that resembled rock, and improved brine recognition persisted across the tested 60-, 180- and 300-s intervals between input images. Because the model uses relative image positions rather than elapsed time in seconds, it could also be trained for more frequent imaging. Applying this approach to faster laboratory CT could broaden the range of pore-scale flow experiments accessible with laboratory systems and reduce reliance on synchrotron facilities.

The transfer tests extend this comparison to mixed-wet and water-wet dissolution sequences from another Ketton sample. When applied to the new sequences without retraining, the 4D models retained higher phase Dice and most boundary F1 scores (Table~\ref{tab:dissolution_transfer}). Temporal information continued to support the identification of rock and fluid regions as the experimental conditions changed. This extends the segmentation comparison to related dissolution experiments with different wettability and imaging conditions.

Our results show that segmentation errors can distort measurements of pore-space evolution and fluid displacement and alter the physical properties calculated from images. Even when porosity is unchanged, differences in pore-boundary geometry can change calculated permeability and mineral dissolution rate. Segmentation should therefore be evaluated through the physical measurements and simulations it supports, alongside phase-identification accuracy. By using information from other images in the time sequence, PoreTimeNet improves measurements of pore-volume growth and oil filling and estimates of permeability changes. Extending this approach to faster laboratory CT could help measure changes in pore geometry and fluid occupancy at higher temporal resolution. These measurements could help explain how pore-scale processes govern fluid transport in geological storage and groundwater systems.

\section{Methods}
\label{sec2}

\subsection{Data preparation}

We selected time-resolved micro-CT datasets to evaluate whether temporal information improves measurements of pore-space evolution and fluid displacement. Carbonate dissolution allowed us to assess measurements of changing pore geometry. Drainage and imbibition in Ketton limestone tested whether the predicted rock geometry remained consistent as fluids moved through the fixed pore space. Oil flooding in Luxembourg sandstone provided low-contrast laboratory CT images for assessing fluid identification and measurements of oil filling. Two further dissolution sequences from another Ketton sample tested transfer to new experimental conditions without retraining. Table~\ref{tab:dataset_overview} summarises all six sequences.

\begin{table}[!htbp]
\centering
\setlength{\tabcolsep}{3pt}
\renewcommand{\arraystretch}{1.15}
\caption{\textbf{Overview of the time-resolved micro-CT datasets.} Voxel sizes are those used in this study.}
\label{tab:dataset_overview}
\begin{tabular}{@{}p{0.40\textwidth}p{0.25\textwidth}p{0.15\textwidth}l@{}}
\toprule
\raggedright\textbf{Dataset / process} & \raggedright\textbf{Imaging technique} & \raggedright\textbf{Voxel size ($\mu$m)} & \textbf{Data source} \\
\midrule
\raggedright Ketton limestone:\newline water-wet dissolution (Seq1) & \raggedright Laboratory micro-CT & 6.09 & \cite{ma2026time} \\
\addlinespace[3pt]
\raggedright Ketton limestone:\newline mixed-wet dissolution (Seq2) & \raggedright Laboratory micro-CT & 5.2 & \cite{ma2026multiphase} \\
\addlinespace[3pt]
\raggedright Ketton limestone:\newline water-wet dissolution (Seq3) & \raggedright Laboratory micro-CT & 5.2 & \cite{ma2026multiphase} \\
\addlinespace[3pt]
\raggedright Ketton limestone:\newline drainage & \raggedright Synchrotron micro-CT & 3.28 & \cite{singh2018time} \\
\addlinespace[3pt]
\raggedright Ketton limestone:\newline imbibition & \raggedright Synchrotron micro-CT & 3.28 & \cite{singh2018time} \\
\addlinespace[3pt]
\raggedright Luxembourg sandstone:\newline oil flooding & \raggedright Laboratory micro-CT & 8.051 & \cite{mascini2021fluid,mascini2021dataset} \\
\bottomrule
\end{tabular}
\end{table}

\subsubsection{Dissolution of Ketton limestone}
We used ten images from a water-wet Ketton limestone dissolution experiment in which CO$_2$-saturated brine displaced oil~\cite{ma2026time}. Reference segmentations were prepared as described in that study. Mixed-wet Seq2 and water-wet Seq3 were acquired from the same second Ketton sample and used for transfer evaluation without retraining (Table~\ref{tab:dissolution_transfer}; Supplementary Note~\ref{supp-sec:transfer_matched}).

\subsubsection{Drainage and imbibition in Ketton limestone}
The drainage and imbibition sequences reported by Singh et al.~\cite{singh2018time} comprised 496 and 416 images, respectively.

\subsubsection{Luxembourg sandstone}
The Luxembourg sandstone images were acquired continuously, with each reconstruction covering 60\,s~\cite{mascini2021fluid,mascini2021dataset}. The time-step values give the start of each scan relative to the start of continuous imaging. The segmentations provided by the dataset authors for each time step were used as the reference segmentations.

\subsection{Training, validation and evaluation data}
We used separate image regions for training and evaluation. For dissolution (Seq1), training used \(x\in[0,900)\), \(y,z\in[0,1050)\). The validation region occupied \(x\in[900,1000)\), \(y,z\in[275,775)\). Predictions were evaluated within \(x\in[1000,1500)\), \(y,z\in[275,775)\). These are voxel coordinates in the original images.

For Ketton drainage, we evaluated a \(500^3\)-voxel region outside the training region. For imbibition, we excluded the training region and nearby voxels whose predictions could use image data from it. In local voxel coordinates, the evaluated region was \([0,500)^3\setminus[144,500)^3\). Within each Ketton process, all methods were evaluated on the same voxels. The labelled training time steps are listed in the Supplementary Methods.

Reference segmentations for evaluation were available at time-step indices 1, 100, 200, 300 and 400 for drainage, and 1, 100, 170, 200, 230 and 380 for imbibition. Drainage evaluation shared indices 1, 100 and 200 with training supervision, and imbibition evaluation used the same six indices as training supervision; in both processes these time steps were scored in a region spatially separate from the training labels.

For Luxembourg sandstone, all regions shared \(x\in[135,860)\) and \(y\in[139,867)\). The training and validation regions were \(z\in[8,323)\) and \(z\in[323,451)\), respectively. Predictions were evaluated over \(z\in[323,579)\), which included the validation region.

Validation used the separate validation region at 0, 16, 27, 32, 37, 47, 58, 69 and 74\,min. The reference segmentations used for this model selection were additional to those used for training.

Luxembourg sandstone models trained with different numbers of labelled time steps were evaluated at 0, 5, 11, 16, 21, 27, 32, 37, 42, 47, 53, 58, 63, 69 and 74\,min. Headline averages excluded 37\,min, shared by all training-label sets. The reported brine Dice scores and volume errors were calculated using the remaining 14 images.

\subsection{Problem formulation}

Let
\begin{equation}
\mathcal{V}=\{V_i \in \mathbb{R}^{D \times H \times W}\}_{i=1}^{N}
\end{equation}
denote a time-resolved micro-CT sequence, where \(V_i\) is the three-dimensional greyscale image acquired at time step \(i\), and \(N\) is the number of time steps. We assign a phase label
\begin{equation}
y_i(\mathbf{x}) \in \{0,\ldots,K-1\}
\end{equation}
to every voxel \(\mathbf{x}\in\Omega\) at every time step, where \(\Omega\) is the three-dimensional voxel domain and \(K\) is the number of segmentation phases.

Conventional 3D segmentation learns an independent mapping
\begin{equation}
f_{\mathrm{3D}}: V_i \mapsto \hat{y}_i ,
\end{equation}
which segments each 3D image separately, ignoring temporal continuity in displacement and preventing information exchange between images in the time sequence.

We formulate segmentation as joint spatiotemporal prediction. For an ordered temporal window
\begin{equation}
W_m=\{V_{i_1},V_{i_2},\ldots,V_{i_T}\},
\end{equation}
where \(T\) is the number of images processed jointly by the network and \(i_1,\ldots,i_T\) are time-step indices, the model predicts the corresponding label maps:
\begin{equation}
f_{\mathrm{4D}}: W_m \mapsto \{\hat{y}_{i_1},\hat{y}_{i_2},\ldots,\hat{y}_{i_T}\}.
\end{equation}
Here, \(T\) is the number of time steps in the input window. PoreTimeNet can thus use context from other images in the input window while preserving the full three-dimensional pore geometry at each time step.

Training supervision is restricted to the available labelled positions within each temporal window, while predictions cover all \(T\) window positions. The number of labelled time steps and the spatial support of their labels depend on the dataset.

\subsection{PoreTimeNet framework}
\label{sec:architecture}

PoreTimeNet consists of a shared Swin Transformer spatial encoder, multi-scale temporal attention between encoder stages, and a decoder with temporal fusion at its three coarsest stages. Supplementary Fig.~\ref{supp-fig:supp_architecture} details the architecture.

\subsubsection{Spatial encoder}

The temporal dimension of \(X \in \mathbb{R}^{B \times C \times T \times D \times H \times W}\) is folded into the batch axis, giving \(X' \in \mathbb{R}^{(B \cdot T) \times C \times D \times H \times W}\). A shared four-stage 3D Swin Transformer~\cite{liu2021swin} applies window-based multi-head self-attention (W-MSA) within non-overlapping \(7 \times 7 \times 7\) windows and shifted-window attention (SW-MSA) across windows (Supplementary Fig.~\ref{supp-fig:supp_architecture}). Patch merging halves spatial resolution after each stage. Multi-scale channel dimensions are \(\{C_0,2\times C_0,4\times C_0,8\times C_0,16\times C_0\}\), with base embedding dimension \(C_0\).

\subsubsection{Temporal attention module}
\label{sec:temporal_attention}

After each spatial Swin block, \(F \in \mathbb{R}^{B \times C \times T \times D \times H \times W}\) is rearranged into
\begin{equation}
    \tilde{F} \in \mathbb{R}^{(B \cdot D \cdot H \cdot W) \times T \times C}.
\end{equation}
Temporal self-attention (T-SA) independently links corresponding spatial locations across all \(T\) time steps. For head \(h\in\{1,\ldots,N_h\}\) with feature dimension \(d_h\),
\[
Q_h=\tilde{F}W_h^Q,\qquad
K_h=\tilde{F}W_h^K,\qquad
V_h=\tilde{F}W_h^V,
\]
and learnable symmetric bias \(B_{\Delta}\in\mathbb{R}^{T\times N_h}\) uses absolute temporal distance \(\Delta=|i-j|\). Attention is
\begin{equation}
\mathrm{T\text{-}SA}_h(\tilde{F})
=
\mathrm{Softmax}\left(
\frac{Q_hK_h^\top}{\sqrt{d_h}} + B_{|i-j|,h}
\right)V_h ,
\label{eq:temporal_attention}
\end{equation}
with \(B_{|i-j|,h}\) the bias between steps \(i\) and \(j\) in head \(h\). Head outputs are concatenated and linearly projected.

\subsubsection{Temporal fusion decoder}

Sequential UnetrUpBlock modules upsample bottleneck features and fuse encoder skip connections through \(3 \times 3 \times 3\) convolutions. At each of the three coarsest stages, spatial decoding is followed by a pre-normalisation TemporalFusionBlock:
\begin{align}
    F'  &= F + \mathrm{DropPath}\bigl(\mathrm{T\text{-}SA}(\mathrm{LN}(F))\bigr), \label{eq:fusion_attn} \\
    F'' &= F' + \mathrm{DropPath}\bigl(\mathrm{MLP}(\mathrm{LN}(F'))\bigr),        \label{eq:fusion_mlp}
\end{align}
Here, \(\mathrm{LN}\) denotes layer normalisation and \(\mathrm{T\text{-}SA}\) follows Eq.~\eqref{eq:temporal_attention}. These blocks refine predictions using features from other images in the input window. The two finest stages omit temporal fusion to reduce memory use at high spatial resolutions. A \(1 \times 1 \times 1\) convolution produces logits \(\mathbf{z} \in \mathbb{R}^{B \times K \times T \times D \times H \times W}\) for all \(K\) phases.

\subsection{Multi-time-step supervision}
\label{sec:loss}

For each training window, the model predicts the phases in every image. The segmentation loss compares these predictions with reference segmentations wherever training labels are available within the selected training region. For batch sample \(b\) with window \(W_b\), supervised entries \(\mathcal{S}_b(W_b)\) contain pairs \((L_{b,j},q_{b,j})\). Here, \(j\) indexes labelled entries, \(L_{b,j}\) is the voxel label map, \(q_{b,j}\in\{0,\ldots,T-1\}\) its temporal position, and \(v\) a patch voxel position \((d,h,w)\). Class-wise softmax gives
\begin{equation}
p_{b,k,q,v}
=
\frac{\exp(z_{b,k,q,v})}
{\sum_{k'=0}^{K-1}\exp(z_{b,k',q,v})}.
\end{equation}
The training objective was
\begin{equation}
\mathcal{L}
=
\lambda_{\mathrm{Dice}}\mathcal{L}_{\mathrm{Dice}}
+
\lambda_{\mathrm{CE}}\mathcal{L}_{\mathrm{CE}}.
\label{eq:total_loss}
\end{equation}
Dice and cross-entropy losses cover all three phases at supervised positions, averaged over labelled entries and batch elements. Cross-entropy uses label smoothing \(\alpha\) and class weights
\[
\boldsymbol{\omega}=[\omega_{\text{rock}},\,\,\omega_{\text{oil}},\,\omega_{\text{brine}}].
\]

\subsection{Training settings and model comparisons}

Each training input consisted of three-dimensional patches cropped from the same spatial location at several time steps. These patches formed a temporal window. Supplementary Table~\ref{supp-tab:dataset_settings} summarises the training and inference settings.

For each comparison, the 3D and 4D models used the same random seeds: 42, 1234 and 2026. All models were trained from random initialisation. The 3D model used the same spatial encoder and decoder as the 4D model but omitted temporal attention and temporal decoder fusion.

For Ketton drainage and imbibition and Luxembourg sandstone, each input contained 11 image patches of \(96^3\) voxels. Ketton input time steps were separated by 1, 3, 5 or 10 sequence positions, with each interval assigned the same total sampling weight. The sandstone training windows used 60-, 180- or 300-s intervals between input images, sampled equally. Dissolution inputs contained \(128^3\)-voxel patches from five consecutive time steps.

In Luxembourg sandstone, we varied the number of labelled training time steps. Training with one, three or five labelled time steps used reference segmentations at $\{37\}$, $\{18,37,56\}$ and $\{9,18,37,56,65\}$ min, respectively. For each architecture, models were trained separately with each set of labelled time steps using the three seeds listed above and the same training program as the Ketton drainage and imbibition models. When training used one labelled time step, only the 37-min reference segmentation supervised training. The 3D model learned from that image, while the 4D model also used other images in the input window without training labels.

The 3D and 4D models used AdamW with a weight decay of \(10^{-5}\). Models for the displacement datasets were trained for 300 epochs, with a peak learning rate of \(1.4142\times10^{-5}\), a linear increase during 30 warm-up epochs and then cosine decay to \(10^{-6}\). Dissolution models were trained for 250 epochs, with a peak learning rate of \(2\times10^{-5}\), 50 warm-up epochs and cosine decay to \(10^{-6}\) over the remaining 200 epochs. Each GPU processed one window per batch on two GPUs; model parameters were updated after each batch without gradient accumulation. Luxembourg sandstone training used 115 batches per epoch.

For the 3D and 4D models, we evaluated fluid Dice scores in the validation region every ten epochs and retained the saved model, or checkpoint, with the highest score.

Dice and cross-entropy losses had equal weights, with cross-entropy label smoothing \(\alpha\) of 0.05. Rock, oil and brine class weights were \((1,1,1)\) for limestone and sandstone displacement experiments and \((0.8,1.8,1.0)\) for dissolution.

For all datasets, training patches were sampled using class-balanced spatial crops. Data augmentation included axis flips, affine transformations and Gaussian smoothing. The same spatial transformation was applied to all images in a window. Temporal intensity drift, which varied greyscale values across the window, was applied with probability 0.5. Reversing the order of the input images was enabled for the displacement datasets and disabled for dissolution. Input corruption applied noise, masking or intensity jitter to a selected training image. In the masking branch, the same spatial region in other images selected from the input window could also receive weaker perturbations. The same input-corruption parameters were used wherever it was enabled. The Supplementary Methods give the probabilities and parameter ranges.

For dissolution, we made three additional comparisons to test whether the 4D model's improved accuracy came from information in other images in the input window. First, we trained an additional 3D model with input corruption. Second, we trained the current-time-step-only 4D model. It retained the full architecture and its 69,237,335 parameters, but restricted every temporal attention matrix to its diagonal during training and inference. Other time steps could therefore not influence the prediction. Query/key and temporal-position parameters receiving no gradient under this restriction comprised 1,175,186 parameters; 68,062,149 parameters continued to receive gradients. Third, during inference, we repeated the current image five times to form the input sequence, keeping the trained 4D model's weights unchanged. All comparisons used the same full-volume evaluation. The four time steps at which the 3D model's three-seed mean rock boundary F1 score over the complete \(500^3\) region was below 0.80 were 0, 300, 356 and 471\,min; overall comparisons included all ten time steps.

For Ketton drainage and imbibition, input corruption was disabled for the 3D model and enabled for the 4D model. We also trained an additional Ketton 3D model with input corruption. For dissolution, input corruption was disabled for both the 3D and 4D models and for the current-time-step-only 4D model. For Luxembourg sandstone, input corruption was enabled for the 4D models and disabled for the 3D models.

\subsection{Inference and evaluation metrics}\label{sec:training}\label{sec:inference_evaluation}
To segment an image, the 4D model received that image together with other images from the time sequence within a temporal window. For the results in Figs.~\ref{fig:dissolution_measurements}--\ref{fig:pore_dynamics}, we retained only its prediction for the middle image, termed the central time-step output. Within each Ketton drainage or imbibition input window, images were separated by five sequence positions; dissolution and the sandstone windows used consecutive images. Where a window extended beyond the sequence, the nearest endpoint image was repeated. We also evaluated Gaussian averaging of predictions from overlapping temporal windows as a separate Ketton variant and used it for the cross-sample predictions in Table~\ref{tab:dissolution_transfer}, as described below.

Inference used the training patch sizes and, for the 4D model, window lengths. Both models used 50\% spatial overlap and Gaussian blending of class probabilities from overlapping patches.

For each temporal window \(W_m\), the network produced class scores, or logits, at all \(T\) positions. Within each spatial patch, class-wise softmax converted these scores to probabilities. We combined probabilities from overlapping spatial patches using Gaussian weights and divided by the sum of those weights. The result, \(p^{(m)}_{i,k}(\mathbf{x})\), is the probability for class \(k\) at voxel \(\mathbf{x}\) of time step \(i\). When using the central time-step output, each voxel was assigned to the phase with the highest probability. When endpoint padding repeated a time step within a window, only the output closest to the window centre was retained for that time step.

The temporal-aggregation variant combined predictions for the same image obtained from different temporal windows. Let \(\mathcal{M}_i\) index windows containing time step \(i\), and \(c_m\) denote the central time-step index of window \(m\). The weight assigned to each prediction was
\begin{equation}
\begin{aligned}
g_{m,i} &= \exp\!\left[-\frac{(i-c_m)^2}{2\sigma_{\mathrm{temp}}^2}\right], \\
w_{m,i} &= g_{m,i}\,\mathbb{I}(g_{m,i}\geq\eta),
\end{aligned}
\label{eq:inference_temporal_weights}
\end{equation}
where \(\sigma_{\mathrm{temp}}>0\) is the Gaussian width in time-step-index intervals and \(0\leq\eta\leq1\) the minimum retained weight. The indicator \(\mathbb{I}(\cdot)\) equals one for a true argument and zero otherwise. For this additional comparison, drainage and imbibition used \(\sigma_{\mathrm{temp}}=5\) time-step-index intervals and \(\eta=0.05\). The distance \(i-c_m\) counts positions in the image sequence. The weight threshold determines which predictions are averaged; all \(T\) images are still supplied to the network.

For the separate temporal-aggregation variant, the class probabilities and segmentation were
\begin{equation}
\begin{aligned}
\bar p_{i,k}(\mathbf{x})
&=\frac{\displaystyle\sum_{m\in\mathcal{M}_i}w_{m,i}\,p^{(m)}_{i,k}(\mathbf{x})}
{\displaystyle\sum_{m\in\mathcal{M}_i}w_{m,i}}, \\
\hat y_i(\mathbf{x})
&=\operatorname*{arg\,max}_{k\in\{0,\ldots,K-1\}}\bar p_{i,k}(\mathbf{x}).
\end{aligned}
\label{eq:inference_temporal_aggregation}
\end{equation}
Each evaluated time step had a centred window with unit weight, ensuring a positive denominator. All terms predict the same time step \(i\) from different window contexts. This output aggregation is separate from learned temporal feature fusion.

The 3D model used MONAI sliding-window inference~\cite{cardoso2022monai}. Each 3D prediction used only the image \(V_i\).

At each time step with a reference segmentation, \(Y_k\) and \(\widehat Y_k\) denote reference and predicted voxel sets for class \(k\), restricted to the evaluation region and valid sample support. We assessed both phase overlap and boundary agreement~\cite{maierhein2024metrics}. The class-specific Dice coefficient~\cite{dice1945measures} was
\begin{equation}
D_k=\frac{2\,|Y_k\cap\widehat Y_k|}{|Y_k|+|\widehat Y_k|},
\label{eq:evaluation_dice}
\end{equation}
where \(|\cdot|\) counts voxels.

The boundary F1 score used a three-dimensional adaptation of the class-specific BF score~\cite{csurka2013good}. The inner boundary and its matching neighbourhood were
\begin{equation}
\partial Y_k=\{\mathbf x\in Y_k:\mathcal N_6(\mathbf x)\not\subseteq Y_k\},\qquad
B(S)=\{\mathbf x:\min_{\mathbf z\in S}\lVert\mathbf x-\mathbf z\rVert_2\le 1\},
\label{eq:boundary_sets}
\end{equation}
where \(\mathcal N_6(\mathbf x)\) contains the six face neighbours of \(\mathbf x\); the same boundary definition applies to \(\widehat Y_k\). Voxels outside the evaluation region, valid support or array belong to no class. The boundary consists of voxels removed by one six-neighbour binary erosion. Distances are measured in voxel units: \(d\le1\) accepts the same voxel or a face neighbour, but not an edge or corner neighbour. Boundary precision, recall and F1 scores were
\begin{equation}
\mathrm{Prec}_k=\frac{|\partial\widehat Y_k\cap B(\partial Y_k)|}{|\partial\widehat Y_k|},\quad
\mathrm{Rec}_k=\frac{|\partial Y_k\cap B(\partial\widehat Y_k)|}{|\partial Y_k|},\quad
\mathrm{BF1}_k=\frac{2\,\mathrm{Prec}_k\,\mathrm{Rec}_k}{\mathrm{Prec}_k+\mathrm{Rec}_k}.
\label{eq:boundary_f1}
\end{equation}
Each boundary voxel counts once; multiple voxels may match the same counterpart. The BF1 score was zero when both precision and recall were zero. The Supplementary Methods give the empty-set conventions.

The fluid Dice score was \(D_{\mathrm{fluid}}=(D_{\mathrm{oil}}+D_{\mathrm{brine}})/2\). The fluid boundary F1 score likewise averaged the oil and brine scores. The three-class mean Dice score equally weighted rock, oil and brine.

\subsection{Porosity, fluid saturation and temporal consistency}
For oil, brine and rock voxel counts \(N_o,N_b,N_r\) within the valid region, oil saturation and porosity were
\begin{equation}
S_o=\frac{N_o}{N_o+N_b},\qquad
\phi=\frac{N_o+N_b}{N_o+N_b+N_r}.
\label{eq:transfer_bulk_metrics}
\end{equation}
Absolute differences from the reference values are reported in percentage points. Porosity and saturation were computed from each segmentation's own phase volumes. For the sandstone dataset, brine-volume error was the absolute relative difference from the reference brine volume.

Sequence diagnostics measured changes in predicted rock within fixed spatial support. For the displacement datasets, changes in predicted rock were measured as the percentage of valid voxels whose predictions changed between rock and pore in consecutive images. Mean absolute porosity change was the mean of the absolute porosity differences over the same image pairs. Ketton sequence summaries used time steps 26--471 for drainage and 26--391 for imbibition, excluding 25 time steps at each end. For the Ketton repeated-input comparison, changes in predicted rock were evaluated over time steps 221--278 for drainage and 181--238 for imbibition (Supplementary Table~\ref{supp-tab:ketton_repeated}). For dissolution, we counted pore-to-rock transitions between consecutive time steps; three-time-step voting of the predictions was assessed alongside its effect on accuracy against the reference segmentations (Supplementary Note~\ref{supp-sec:d2_consistency}).

\subsection{Statistics and reproducibility}

For time-averaged metrics, values were averaged across the designated time steps within each training run, then summarised as means and sample standard deviations across three seeds. Comparisons were paired by seed. Statistical replication was defined by the training seed, with all models trained on the same rock sample.

\subsection{Segmentation-derived flow and transport simulations}
\label{sec:geometry_flow_transport}
We used GeoChemFoam 5.2 (OpenFOAM v2212)~\cite{maes2021geochemfoam} to calculate flow and reactive transport in the complete \(500^3\)-voxel dissolution region at 0, 187 and 646\,min. Each reference or predicted segmentation defined a separate fixed geometry. Oil and brine voxels were combined into pore space, and only face-connected pore clusters spanning the inlet and outlet were retained. Each retained voxel became one hexahedral cell of side \(6.09\,\mu\mathrm{m}\). The calculations filled this geometry with a single fluid to measure the effect of segmentation on intrinsic permeability.

Steady incompressible flow used an inlet flow rate of \(Q=5.4655\times10^{-10}\,\mathrm{m^3\,s^{-1}}\), obtained by scaling the experimental injection rate of \(0.1\,\mathrm{mL\,min^{-1}}\) to the cross-sectional area of the analysed region. The outlet pressure was zero, solid walls were no-slip, and the kinematic viscosity was \(\nu=6.1194\times10^{-7}\,\mathrm{m^2\,s^{-1}}\). Permeability was calculated as
\begin{equation}
K=\frac{\nu Q/A}{|\mathrm{d}\langle p_{\mathrm{kin}}\rangle/\mathrm{d}x|},
\end{equation}
where \(A=(500h)^2\) is the total cross-sectional area, \(h\) is the voxel size, and \(\langle p_{\mathrm{kin}}\rangle\) is the mean kinematic pressure over fluid cells in each plane normal to the flow. We fitted the pressure gradient by least squares over planes 100--399, while solving flow in the full geometry. Pressure values plotted in pascals use a fluid density of \(1340\,\mathrm{kg\,m^{-3}}\).

Steady transport was then solved on the fixed velocity field with diffusivity \(7.5\times10^{-10}\,\mathrm{m^2\,s^{-1}}\). The transported concentration represented the effective remaining capacity of the brine to dissolve calcite, with an effective inlet value of \(25\,\mathrm{mol\,m^{-3}}\). Rock walls used a first-order reactive condition with coefficient \(k=10^{-6}\,\mathrm{m\,s^{-1}}\) and unit stoichiometric coefficient. Lateral domain boundaries were non-reactive, with zero normal concentration gradient; the outlet also had zero concentration gradient. The total mineral dissolution rate was
\begin{equation}
R=k\int_{\Gamma_{\mathrm{r}}}C\,\mathrm{d}A,
\end{equation}
where \(\Gamma_{\mathrm{r}}\) denotes the reactive rock walls and \(C\) is in \(\mathrm{mol\,m^{-3}}\). Each geometry was held fixed during the calculation. All cases used the same physical parameters, boundary conditions and native voxel spacing. Reported errors were relative to the calculation on the reference segmentation at the same time step.

The flow and transport calculations included three reference geometries and predictions from the 3D model, the additional 3D model trained with input corruption, and the 4D model. Each model contributed predictions from three training seeds at each time step, giving 30 cases. All met flow criteria of pressure residual \(\leq5\times10^{-7}\), each velocity-component residual \(\leq2\times10^{-5}\), and relative inlet/outlet flow error \(\leq10^{-6}\). Transport acceptance combined a concentration-residual target or a stable-rate criterion with concentration and mass-balance checks. Random perturbations of reference pore boundaries tested the effect of the spatial arrangement of segmentation errors: pore volume was increased or decreased by 1, 3 or 5\%, or held constant by equal changes in both directions. These perturbation calculations and the fields used for visualisation included calculations with different numerical tolerances. Their settings, checks and the random-sampling procedure are given in Supplementary Methods; Supplementary Table~\ref{supp-tab:flow_visualisation_settings} lists the flow-solver settings for the visualised fields.

\section*{Data availability}
The Ketton drainage and imbibition data are available from the British Geological Survey National Geoscience Data Centre~\cite{singh2018time}. Luxembourg sandstone data are available from the Digital Rocks Portal~\cite{mascini2021dataset}. The dissolution data are available from Ma et al.~\cite{ma2026time,ma2026multiphase}.

\section*{Code availability}
The PoreTimeNet code is available at \url{https://github.com/ImperialCollegeLondon/PoreTimeNet}. SwinUNETR is available through MONAI (\url{https://github.com/Project-MONAI/MONAI}), and nnU-Net is available at \url{https://github.com/MIC-DKFZ/nnUNet}. The selected trained model weights are available on Zenodo~\cite{ma2026poretimenetweights}. Flow and transport calculations use GeoChemFoam~\cite{maes2021geochemfoam}.

\backmatter
\section*{Acknowledgements}
We acknowledge computational resources and support provided by the Imperial College Research Computing Service (\url{https://doi.org/10.14469/hpc/2232}). This work was supported by the Resource Geophysics Academy (RGA) at Imperial College London.

\section*{Author contributions}
Z.M., B.B. and M.J.B. conceived the study. Z.M. developed the code and wrote the manuscript. Z.M. and R.C. performed the analyses. Z.M. and Q.M. collected and prepared the data. Y.W., B.B. and M.J.B. revised and edited the manuscript.

\section*{Competing interests}
The authors declare no competing interests.

\clearpage
\newgeometry{left=18mm,right=18mm,top=20mm,bottom=20mm,bindingoffset=0mm}
\pdfbookmark[0]{Supplementary Information}{supplementary-information}
\begingroup
\makeatletter
\def\ps@plain{\let\@oddhead\@empty\let\@evenhead\@empty\def\@oddfoot{\hfil\thepage\hfil}\let\@evenfoot\@oddfoot}
\pagestyle{plain}
\renewcommand\normalsize{\@setfontsize\normalsize{12pt}{14.5pt}}
\renewcommand\small{\@setfontsize\small{11pt}{13.6pt}}
\renewcommand\footnotesize{\@setfontsize\footnotesize{10pt}{12pt}}
\let\table\tableorg
\let\endtable\endtableorg
\let\@currentlabel\@empty
\makeatother
\normalsize
\setcounter{secnumdepth}{0}
\setcounter{section}{0}
\setcounter{subsection}{0}
\setcounter{figure}{0}
\setcounter{table}{0}
\setcounter{equation}{0}
\setcounter{suppnote}{0}
\renewcommand{\theHsection}{supp.\arabic{section}}
\renewcommand{\theHsubsection}{supp.\arabic{section}.\arabic{subsection}}
\renewcommand{\theHfigure}{supp.\arabic{figure}}
\renewcommand{\theHtable}{supp.\arabic{table}}
\renewcommand{\theHequation}{supp.\arabic{equation}}
\providecommand{\theHsuppnote}{supp.\arabic{suppnote}}
\renewcommand{\figurename}{Supplementary Fig.}
\renewcommand{\tablename}{Supplementary Table}
\setlength{\parindent}{0pt}
\setlength{\parskip}{0.5\baselineskip}
\setlength{\tabcolsep}{2.75pt}
\renewcommand{\arraystretch}{1.13}
\captionsetup{font=normalsize,labelfont=bf,labelsep=period,justification=raggedright,singlelinecheck=false,skip=8pt}
\captionsetup[table]{position=top}
\raggedbottom

\begingroup\thispagestyle{empty}
\centering
{\Large Supplementary Information\par}
\vspace{1cm}
{\LARGE Spatiotemporal learning improves quantification of pore-space evolution and fluid displacement\par}
\vspace{0.8cm}
Zhuangzhuang Ma\textsuperscript{1,2},
Rukuan Chai\textsuperscript{2,*},
Qianqian Ma\textsuperscript{1,2},
Yanghua Wang\textsuperscript{1,2},
Branko Bijeljic\textsuperscript{2},
Martin J. Blunt\textsuperscript{2,*}\par
\vspace{1cm}
\textsuperscript{1}Resource Geophysics Academy, Imperial College London, London, SW7 2BP, United Kingdom\par
\vspace{0.5cm}
\textsuperscript{2}Department of Earth Science and Engineering, Imperial College London, London, SW7 2AZ, United Kingdom\par
\vspace{0.8cm}
\textsuperscript{*}Corresponding authors:\par
Rukuan Chai: \href{mailto:r.chai@imperial.ac.uk}{r.chai@imperial.ac.uk}\par
Martin J. Blunt: \href{mailto:m.blunt@imperial.ac.uk}{m.blunt@imperial.ac.uk}
\vfill
\begin{flushleft}
\textbf{Contents}\par
\hyperref[supp-sec:si_figures]{Supplementary Figures}\hfill\pageref{supp-sec:si_figures}\par
\hyperref[supp-sec:si_tables]{Supplementary Tables}\hfill\pageref{supp-sec:si_tables}\par
\hyperref[supp-sec:si_notes]{Supplementary Notes}\hfill\pageref{supp-sec:si_notes}\par
\hyperref[supp-sec:si_methods]{Supplementary Methods}\hfill\pageref{supp-sec:si_methods}\par
\hyperref[supp-sec:si_references]{Supplementary References}\hfill\pageref{supp-sec:si_references}
\end{flushleft}
\par\endgroup\clearpage

\section{Supplementary Figures}\label{supp-sec:si_figures}
\begin{figure}[H]
\centering
\includegraphics[width=\textwidth,height=0.53\textheight,keepaspectratio]{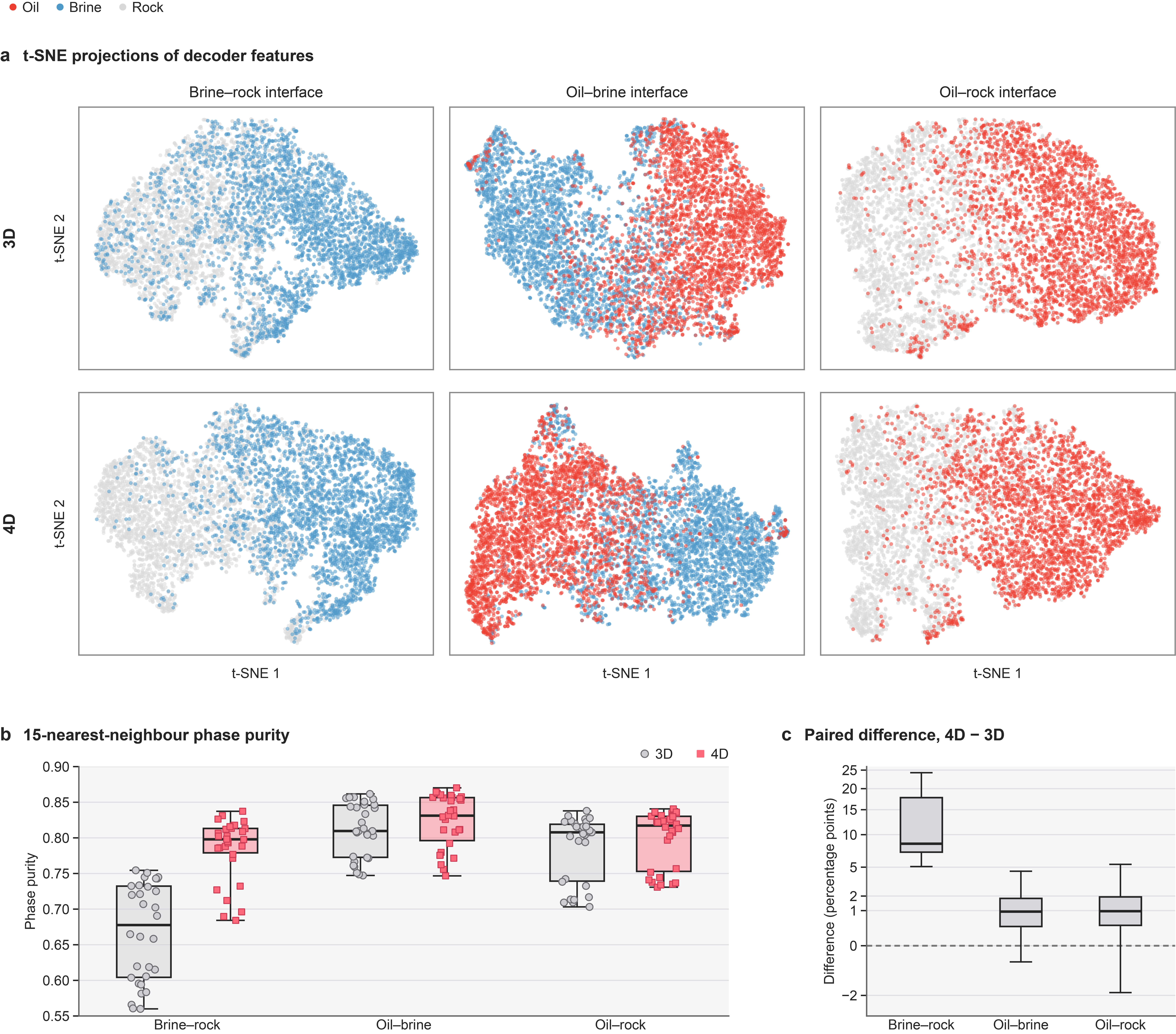}
\caption{\textbf{Network features at phase boundaries during dissolution.} Samples cover the complete $500^3$ Seq1 evaluation region, with identical coordinates for 3D and 4D. Features were extracted from $628^3$ inputs using $128^3$ patches, 50\% overlap and Gaussian weighting ($\sigma=16$ voxels), then L2-normalised. \textbf{a}, t-SNE projections for training seed 42, coloured by reference phase. Projections were fitted independently, so their coordinate axes cannot be compared. \textbf{b}, Phase purity in the original 48-dimensional feature space, with one point per time step and training seed (ten time steps and three seeds). \textbf{c}, Paired 4D-minus-3D differences in percentage points. The vertical axis uses a signed square-root scale, with ticks labelled in the original units. Boxes show the median and interquartile range; whiskers span the minimum to maximum of the 30 values. Mean brine--rock purity increased from 0.668 to 0.785. The paired increase was $11.72\pm0.76$ percentage points, calculated by averaging over time steps within each seed and then taking the mean and sample standard deviation across three seeds.}
\label{supp-fig:decoder_features}
\end{figure}
\clearpage
\begin{figure}[H]
\centering
\includegraphics[width=\textwidth]{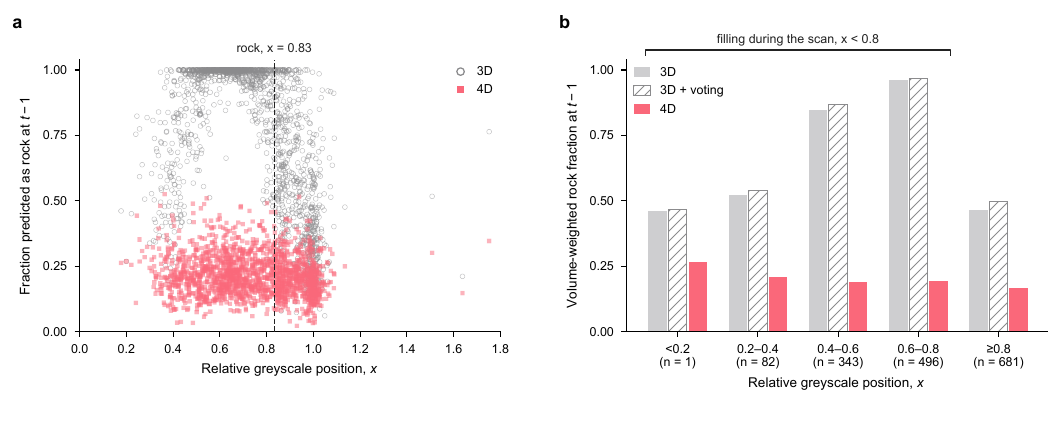}
\caption{\textbf{Predicted phases in pores during oil filling at different greyscale values.} The analysis includes 1,603 fixed reference pores of at least 100 voxels, each with one brine-to-oil transition between neibourghing time steps during 5--69\,min in the combined validation and test region. The selection uses reference segmentations only. \textbf{a}, Fraction of each pore predicted as rock by the 3D model (circles) and the 4D model (squares) in the image preceding the reference transition, plotted against its relative greyscale position $x$. Values near 1 resemble the earlier brine-filled state; values near 0 resemble the later oil-filled state. The dashed line marks the median rock position ($x_{\mathrm{rock}}\approx0.83$). \textbf{b}, Volume-weighted rock fractions in the five greyscale intervals; $n$ gives the number of pores. The bracket marks $x<0.8$, the criterion used to identify filling during the scan. For 3D and 4D, values average training seeds 42, 1234 and 2026. Temporal voting uses 3D seed 42, with two preceding and two following time steps around the image being classified, including the central image. Definitions and per-seed results are given in Supplementary Note~\ref{supp-sec:lux_midscan_filling} and Table~\ref{supp-tab:lux_midscan_rock}.}
\label{supp-fig:lux_midscan_filling}
\end{figure}
\clearpage
\begin{figure}[H]
    \centering
    \includegraphics[width=1.0\textwidth]{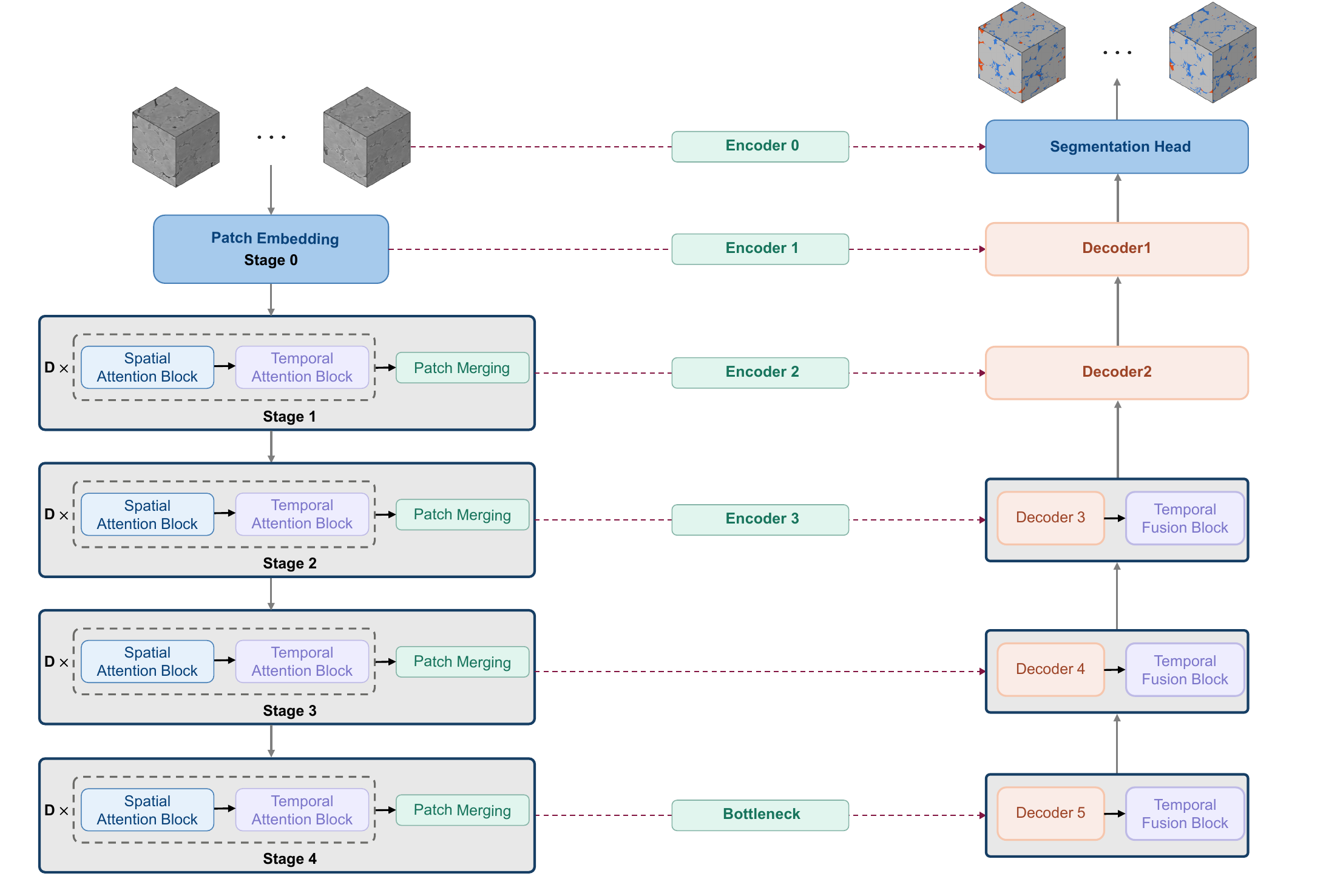}
    \caption{\textbf{Detailed PoreTimeNet architecture.}
    The model receives a temporal window of 3D images as input and predicts phase logits for all time steps within the window. A shared Swin-Transformer encoder extracts hierarchical spatial features from each 3D image, after which temporal self-attention propagates information across corresponding spatial locations in the temporal window. The decoder uses skip connections and decoder-stage temporal fusion to combine multi-scale spatial and temporal features before projecting the output to oil, brine and rock logits.}
    \label{supp-fig:supp_architecture}
\end{figure}
\clearpage
\section{Supplementary Tables}\label{supp-sec:si_tables}
\begin{table}[H]
\centering

\caption{\textbf{Segmentation accuracy during dissolution.} The boundary F1 (BF1) score measures agreement with reference boundaries at Euclidean distances $d\le1$ voxel. $|\Delta\phi|$ and $|\Delta S_o|$ denote absolute porosity and oil-saturation errors in percentage points (pp). Values are means $\pm$ sample standard deviations across three training seeds after averaging within each seed. Fluid Dice and BF1 scores are equally-weighted means over oil and brine. The current-time-step-only 4D model processed each image independently during training and inference.}
\label{supp-tab:dissolution_temporal_ablation}
\begin{tabular}{lcccc}
\toprule
Metric & 3D & \shortstack{3D + input\\corruption} & \shortstack{Current-time-step-only\\4D} & 4D \\
\midrule
Rock Dice score & \(0.9833\pm0.0002\) & \(0.9831\pm0.0005\) & \(0.9830\pm0.0002\) & \(0.9910\pm0.0004\) \\
Fluid Dice score& \(0.9359\pm0.0008\) & \(0.9355\pm0.0016\) & \(0.9347\pm0.0002\) & \(0.9573\pm0.0016\) \\
Rock BF1 score& \(0.8334\pm0.0028\) & \(0.8306\pm0.0058\) & \(0.8294\pm0.0024\) & \(0.9408\pm0.0052\) \\
Fluid BF1 score& \(0.8594\pm0.0023\) & \(0.8573\pm0.0045\) & \(0.8567\pm0.0009\) & \(0.9313\pm0.0046\) \\
\(|\Delta\phi|\) (pp) & \(0.9667\pm0.0991\) & \(0.9696\pm0.1118\) & \(1.0831\pm0.0876\) & \(0.2786\pm0.0690\) \\
\(|\Delta S_o|\) (pp) & \(1.3549\pm0.1542\) & \(1.2151\pm0.3309\) & \(1.5032\pm0.0577\) & \(0.9411\pm0.1924\) \\
\bottomrule
\end{tabular}
\end{table}

\begin{table}[H]
\centering

\caption{\textbf{Pore-to-rock changes in predictions and segmentation accuracy.} Pore-to-rock values are percentages of voxels in the complete evaluation region per interval. Values are means $\pm$ sample standard deviations across three training seeds after averaging within each seed. The fluid Dice score is the equally-weighted mean of oil and brine Dice scores. Voting uses three consecutive time steps; current-time-step-only 4D and Repeated-input 4D are defined in main-text Fig.~\ref{fig:input_modes}a.}
\label{supp-tab:d2_consistency}
\begin{tabular}{lcc}
\toprule
Model & Pore-to-rock (\%) & Fluid Dice score\\
\midrule
3D & \(1.6819\pm0.0170\) & \(0.9359\pm0.0008\) \\
3D + input corruption & \(1.6766\pm0.0335\) & \(0.9355\pm0.0016\) \\
3D + voting & \(0.5549\pm0.0095\) & \(0.8990\pm0.0007\) \\
4D & \(1.0441\pm0.0103\) & \(0.9573\pm0.0016\) \\
Repeated-input 4D & \(2.0691\pm0.0342\) & \(0.9216\pm0.0025\) \\
Current-time-step-only 4D & \(1.6653\pm0.0124\) & \(0.9347\pm0.0002\) \\
4D + voting & \(0.6000\pm0.0097\) & \(0.9204\pm0.0014\) \\
\bottomrule
\end{tabular}
\end{table}

\begin{table}[H]
\centering

\caption{\textbf{Segmentation accuracy and consistency during Ketton drainage.} Changes in predicted rock between consecutive time steps were averaged over time steps 26--471, excluding 25 time steps at each end of the sequence. Mean absolute porosity change is the mean of the absolute differences between consecutive time steps. Values are means $\pm$ sample standard deviations across three training seeds. $|\Delta\phi|$ and $|\Delta S_o|$ denote absolute porosity and oil-saturation errors in percentage points (pp). BF1 denotes the boundary F1 score with a one-voxel matching tolerance. Fluid Dice and Fluid BF1 scores average the corresponding oil and brine scores equally.}
\label{supp-tab:ketton_drainage}
\begin{tabular}{lccc}
\toprule
Metric & 3D & 3D + input corruption & 4D \\
\midrule
Rock Dice score& \(0.9938\pm0.0002\) & \(0.9940\pm0.0001\) & \(0.9941\pm0.0002\) \\
Fluid Dice score& \(0.9065\pm0.0001\) & \(0.9080\pm0.0001\) & \(0.9088\pm0.0008\) \\
Rock BF1 score& \(0.9352\pm0.0043\) & \(0.9415\pm0.0022\) & \(0.9422\pm0.0037\) \\
Fluid BF1 score& \(0.9219\pm0.0004\) & \(0.9232\pm0.0014\) & \(0.9247\pm0.0008\) \\
\(|\Delta\phi|\) (pp) & \(0.3194\pm0.1160\) & \(0.3114\pm0.0842\) & \(0.3172\pm0.0886\) \\
\(|\Delta S_o|\) (pp) & \(0.6955\pm0.4100\) & \(0.7189\pm0.3462\) & \(0.7027\pm0.2033\) \\
\shortstack[l]{Changes in predicted\\rock (\%)} & \(0.5898\pm0.0123\) & \(0.5541\pm0.0105\) & \(0.4089\pm0.0205\) \\
Mean absolute porosity change (pp) & \(0.0308\pm0.0027\) & \(0.0317\pm0.0040\) & \(0.0109\pm0.0006\) \\
\bottomrule
\end{tabular}
\end{table}

\begin{table}[H]
\centering

\caption{\textbf{Segmentation accuracy and consistency during Ketton imbibition.} Changes in predicted rock between consecutive time steps were averaged over time steps 26--391, excluding 25 time steps at each end of the sequence. Mean absolute porosity change is the mean of the absolute differences between consecutive time steps. Values are means $\pm$ sample standard deviations across three training seeds. $|\Delta\phi|$ and $|\Delta S_o|$ denote absolute porosity and oil-saturation errors in percentage points (pp). BF1 denotes the boundary F1 score with a one-voxel matching tolerance. Fluid Dice and Fluid BF1 scores average the corresponding oil and brine scores equally.}
\label{supp-tab:ketton_imbibition}
\begin{tabular}{lccc}
\toprule
Metric & 3D & 3D + input corruption & 4D \\
\midrule
Rock Dice score& \(0.9953\pm0.0002\) & \(0.9954\pm0.000025\) & \(0.9957\pm0.0001\) \\
Fluid Dice score& \(0.9442\pm0.0013\) & \(0.9455\pm0.0002\) & \(0.9510\pm0.0006\) \\
Rock BF1 score& \(0.9810\pm0.0009\) & \(0.9805\pm0.0003\) & \(0.9842\pm0.0003\) \\
Fluid BF1 score& \(0.9490\pm0.0021\) & \(0.9499\pm0.0005\) & \(0.9567\pm0.0008\) \\
\(|\Delta\phi|\) (pp) & \(0.1511\pm0.0446\) & \(0.1843\pm0.0252\) & \(0.1360\pm0.0108\) \\
\(|\Delta S_o|\) (pp) & \(0.7096\pm0.1508\) & \(0.5427\pm0.0556\) & \(0.3263\pm0.1205\) \\
\shortstack[l]{Changes in predicted\\rock (\%)} & \(0.4872\pm0.0106\) & \(0.4699\pm0.0012\) & \(0.3183\pm0.0076\) \\
Mean absolute porosity change (pp) & \(0.0109\pm0.0007\) & \(0.0109\pm0.0012\) & \(0.0049\pm0.0002\) \\
\bottomrule
\end{tabular}
\end{table}

\begin{table}[H]
\centering

\caption{\textbf{Ketton segmentation with repeated images.} The fluid Dice score uses the same evaluation regions as the preceding drainage and imbibition tables. For the comparison of changes in predicted rock, we used predictions at time steps 219--278 for drainage and 179--238 for imbibition, and evaluated changes at time steps 221--278 and 181--238, respectively (58 scored time steps per process, after two inputs initialise consecutive-transition metrics). Changes in predicted rock voxels are quantified as the percentage of evaluated voxels that switch between rock and pore in consecutive predictions, averaged over these time steps. The same voxels were evaluated at all time steps within each process. Values are means $\pm$ sample standard deviations across three training seeds; these rates describe the selected time interval. The fluid Dice score is the equally-weighted mean of oil and brine Dice scores.}
\label{supp-tab:ketton_repeated}
\begin{tabular}{lcccc}
\toprule
Process & Metric & 3D & 4D & Repeated-input 4D \\
\midrule
Drainage & Fluid Dice score& \(0.9065\pm0.0001\) & \(0.9088\pm0.0008\) & \(0.9082\pm0.0001\) \\
Drainage & \shortstack[l]{Changes in predicted\\rock (\%)} & \(0.5448\pm0.0136\) & \(0.3758\pm0.0182\) & \(0.5140\pm0.0063\) \\
Imbibition & Fluid Dice score& \(0.9442\pm0.0013\) & \(0.9510\pm0.0006\) & \(0.9494\pm0.0006\) \\
Imbibition & \shortstack[l]{Changes in predicted\\rock voxels (\%)} & \(0.4904\pm0.0105\) & \(0.3214\pm0.0081\) & \(0.4828\pm0.0043\) \\
\bottomrule
\end{tabular}
\end{table}

\begin{table}[H]
\centering

\caption{\textbf{Effect of the number of labelled training time steps on segmentation performance.} The brine Dice score uses the 14 evaluation images remaining after excluding the image at 37\,min. Values are means $\pm$ sample standard deviations across three training seeds after averaging within each seed. Evaluation times are listed in Supplementary Note~\ref{supp-sec:lux_budget}.}
\label{supp-tab:lux_budget}
\begin{tabular}{llc}
\toprule
\shortstack{Number of labelled\\training time steps} & Model & Brine Dice score \\
\midrule
1 & 3D & \(0.6388\pm0.0018\) \\
1 & 4D & \(0.7127\pm0.0019\) \\
3 & 3D & \(0.6458\pm0.0012\) \\
3 & 4D & \(0.7056\pm0.0108\) \\
5 & 3D & \(0.6441\pm0.0009\) \\
5 & 4D & \(0.7088\pm0.0066\) \\
\bottomrule
\end{tabular}
\end{table}

\begin{table}[H]
\centering

\caption{\textbf{Errors in porosity, oil saturation and brine volume.} Brine error is the absolute relative volume error in percent. Porosity and oil-saturation errors are mean absolute errors in percentage points (pp), averaged over the retained evaluation times. Oil saturation was calculated using the pore volume in each segmentation. Values are means $\pm$ sample standard deviations across three training seeds after averaging within each seed. $|\Delta\phi|$ and $|\Delta S_o|$ denote the porosity and oil-saturation errors.}
\label{supp-tab:lux_budget_volume}
\begin{tabular}{llccc}
\toprule
\shortstack{Number of labelled\\training time steps} & Model & \shortstack{Brine-volume\\error (\%)} & \(|\Delta\phi|\) (pp) & \(|\Delta S_o|\) (pp) \\
\midrule
1 & 3D & \(12.63\pm1.13\) & \(0.68\pm0.09\) & \(2.91\pm0.08\) \\
1 & 4D & \(11.56\pm1.31\) & \(0.65\pm0.07\) & \(2.44\pm0.34\) \\
3 & 3D & \(11.40\pm3.22\) & \(0.63\pm0.21\) & \(2.53\pm0.53\) \\
3 & 4D & \(8.73\pm6.25\) & \(0.53\pm0.45\) & \(1.59\pm0.62\) \\
5 & 3D & \(12.85\pm1.75\) & \(0.76\pm0.14\) & \(2.50\pm0.12\) \\
5 & 4D & \(10.12\pm1.46\) & \(0.59\pm0.11\) & \(2.00\pm0.18\) \\
\bottomrule
\end{tabular}
\end{table}

\begin{table}[H]
\centering

\caption{\textbf{Segmentation accuracy at different time intervals between input images.} The fluid Dice score is the equally-weighted mean of oil and brine Dice scores, calculated within each seed. Scores use the same 11 evaluation time points listed in Supplementary Note~\ref{supp-sec:lux_input_spacing}. Values are means $\pm$ sample standard deviations across three training seeds after averaging over time within each seed. The 3D predictions are reused at every input interval. The rows with one labelled training time step correspond to main-text Fig.~\ref{fig:pore_dynamics}c.}
\label{supp-tab:lux_input_spacing}
\begin{tabular}{llccc}
\toprule
\shortstack{Number of labelled\\training time steps} & Model and interval & \shortstack{Brine Dice\\score} & \shortstack{Oil Dice\\score} & \shortstack{Fluid Dice\\score}\\
\midrule
1 & 3D & $0.6409\pm0.0018$ & $0.7766\pm0.0133$ & $0.7087\pm0.0068$ \\
1 & 4D, 60 s & $0.7258\pm0.0027$ & $0.8570\pm0.0049$ & $0.7914\pm0.0016$ \\
1 & 4D, 180 s & $0.7340\pm0.0023$ & $0.8577\pm0.0034$ & $0.7958\pm0.0013$ \\
1 & 4D, 300 s & $0.7402\pm0.0018$ & $0.8631\pm0.0040$ & $0.8017\pm0.0012$ \\
3 & 3D & $0.6477\pm0.0012$ & $0.7265\pm0.0200$ & $0.6871\pm0.0102$ \\
3 & 4D, 60 s & $0.7159\pm0.0112$ & $0.8586\pm0.0019$ & $0.7873\pm0.0058$ \\
3 & 4D, 180 s & $0.7235\pm0.0103$ & $0.8586\pm0.0045$ & $0.7910\pm0.0065$ \\
3 & 4D, 300 s & $0.7295\pm0.0094$ & $0.8571\pm0.0068$ & $0.7933\pm0.0067$ \\
5 & 3D & $0.6459\pm0.0009$ & $0.7416\pm0.0053$ & $0.6938\pm0.0025$ \\
5 & 4D, 60 s & $0.7199\pm0.0070$ & $0.8473\pm0.0080$ & $0.7836\pm0.0043$ \\
5 & 4D, 180 s & $0.7274\pm0.0068$ & $0.8501\pm0.0073$ & $0.7887\pm0.0060$ \\
5 & 4D, 300 s & $0.7332\pm0.0063$ & $0.8474\pm0.0069$ & $0.7903\pm0.0056$ \\
\bottomrule
\end{tabular}
\end{table}

\begin{table}[H]
\centering

\caption{\textbf{Oil-filling volume across the sampled intervals.} For the Luxembourg sandstone models trained with one labelled time step, filling denotes brine-to-oil transitions over the sampled intervals (Supplementary Note~\ref{supp-sec:lux_observed_filling}). Model values are means $\pm$ sample standard deviations across three training seeds; reference volume is common to all comparisons. Signed difference is $100(V_{\mathrm{model}}-V_{\mathrm{reference}})/V_{\mathrm{reference}}$.}
\label{supp-tab:lux_filling_volume}
\begin{tabular}{lcc}
\toprule
Segmentation & Volume (mm$^3$) & Difference (\%) \\
\midrule
Reference segmentation & \(1.942\) & --- \\
3D & \(1.095\pm0.009\) & \(-43.61\pm0.46\) \\
4D & \(1.658\pm0.002\) & \(-14.62\pm0.13\) \\
\bottomrule
\end{tabular}
\end{table}

\begin{table}[H]
\centering
\caption{\textbf{Pore volume predicted as rock before reference oil filling.} Values are volume-weighted rock fractions at $t-1$ within each pore subset. The 3D and 4D columns show individual training seeds. Temporal voting uses only 3D seed 42 and a five-image window centred on $t-1$; it is not a mean over seeds. The relative greyscale position $x$ and the event selection are defined in Supplementary Note~\ref{supp-sec:lux_midscan_filling}.}
\label{supp-tab:lux_midscan_rock}
\small
\begin{tabular}{lrccccccc}
\toprule
 & & \multicolumn{3}{c}{3D} & 3D vote & \multicolumn{3}{c}{4D} \\
Subset & $n$ & 42 & 1234 & 2026 & 42 & 42 & 1234 & 2026 \\
\midrule
All pores & 1603 & 0.708 & 0.697 & 0.695 & 0.721 & 0.172 & 0.187 & 0.181 \\
$x<0.8$ & 922 & 0.909 & 0.891 & 0.893 & 0.908 & 0.185 & 0.200 & 0.191 \\
$0.6\le x<0.8$ & 496 & 0.968 & 0.959 & 0.961 & 0.967 & 0.184 & 0.203 & 0.191 \\
$x\ge0.8$ & 681 & 0.469 & 0.465 & 0.460 & 0.498 & 0.157 & 0.172 & 0.170 \\
\bottomrule
\end{tabular}
\end{table}

\begin{table}[H]

\centering

\caption{\textbf{Comparison with nnU-Net.} Segmentation scores cover all ten time steps and the complete $500^3$ region. For these scores, 3D and 4D values are means $\pm$ sample standard deviations across three training runs after averaging over time within each run; nnU-Net values are time averages from one run. Porosity and saturation errors are averaged after taking their absolute values. Fluid Dice and Fluid BF1 scores average oil and brine equally. The BF1 score uses a one-voxel matching tolerance. $|\Delta\phi|$ and $|\Delta S_o|$ are absolute porosity and oil-saturation errors in percentage points (pp). Permeability ($K$) and total mineral dissolution rate ($R$) are evaluated at the stated times using the same flow and reactive-transport settings as main-text Fig.~\ref{fig:flow_simulation}: $K$ is calculated from the pressure gradient fitted over planes 100--399. For the $K$ and $R$ error rows, 3D and 4D values are means $\pm$ sample standard deviations of the three absolute relative errors, whereas nnU-Net values are signed relative errors from one run; positive and negative signs denote overestimation and underestimation, respectively. The final row reports the permeability ratio, whose reference value is 8.34. No between-run standard deviation is available for nnU-Net.}

\label{supp-tab:nnunet_ct_summary}

\begin{tabular}{lccc}

\toprule

Metric & 3D & nnU-Net & 4D \\

\midrule

Rock Dice score& $0.9833\pm0.0002$ & 0.9800 & $0.9910\pm0.0004$ \\

Fluid Dice score& $0.9359\pm0.0008$ & 0.9223 & $0.9573\pm0.0016$ \\

Rock BF1 score& $0.8334\pm0.0028$ & 0.7920 & $0.9408\pm0.0052$ \\

Fluid BF1 score& $0.8594\pm0.0023$ & 0.8061 & $0.9313\pm0.0046$ \\

\(|\Delta\phi|\) (pp) & $0.97\pm0.10$ & 1.11 & $0.28\pm0.07$ \\

\(|\Delta S_o|\) (pp) & $1.35\pm0.15$ & 1.49 & $0.94\pm0.19$ \\

\midrule
$K$ relative error at 0 min (\%) & $13.0\pm9.6$ & $+30.8$ & $4.6\pm1.2$ \\
$K$ relative error at 187 min (\%) & $6.7\pm3.1$ & $-23.9$ & $0.5\pm0.4$ \\
$K$ relative error at 646 min (\%) & $18.0\pm3.5$ & $-23.4$ & $2.1\pm1.9$ \\
$R$ relative error at 0 min (\%) & $4.1\pm3.4$ & $-0.2$ & $4.5\pm1.2$ \\
$R$ relative error at 187 min (\%) & $3.1\pm0.6$ & $-3.2$ & $1.3\pm0.6$ \\
$R$ relative error at 646 min (\%) & $3.0\pm0.6$ & $-3.4$ & $1.7\pm0.1$ \\
$K(646)/K(0)$ & $6.07\pm0.31$ & $4.88$ & $8.72\pm0.56$ \\

\bottomrule

\end{tabular}

\end{table}

\begin{table}[H]
\centering

\caption{\textbf{Training and inference settings for each dataset.} An input window contains three-dimensional patches from the same spatial region at several time steps. The 3D model processes the images independently; the 4D model combines their information. Intervals below count positions in the image sequence. The models started from random initialisation and used two GPUs with one window per GPU and no gradient accumulation. AdamW weight decay was $10^{-5}$. We retained the checkpoint with the highest validation fluid Dice score. Comparisons with input corruption, current-time-step-only 4D and repeated inputs are described separately.}
\label{supp-tab:dataset_settings}
\begin{tabularx}{\textwidth}{@{}L{0.27\textwidth}YYY@{}}
\toprule
Setting & Dissolution & Ketton\newline displacement & Luxembourg\newline sandstone \\
\midrule
Spatial patch size (voxels) & $128^3$ & $96^3$ & $96^3$ \\
Images per input window & 5 images & 11 images & 11 images \\
Training input intervals & Consecutive time steps & 1, 3, 5, 10 time-step intervals & 1, 3, 5 time-step intervals \\
Epochs / warm-up & 250 / 50 & 300 / 30 & 300 / 30 \\
Peak learning rate & $2\times10^{-5}$ & $1.4142\times10^{-5}$ & $1.4142\times10^{-5}$ \\
Labelled training time steps & 10 & 6 per process & 1, 3 or 5 \\
Class weights (rock, oil, brine) & 0.8, 1.8, 1.0 & 1, 1, 1 & 1, 1, 1 \\
Input corruption (3D / 4D) & Off / off & Off / on & Off / on \\
Input interval at inference & Consecutive time steps & 5 time-step intervals & 1 time-step interval \\
Spatial overlap & 50\% & 50\% & 50\% \\
\bottomrule
\end{tabularx}
\end{table}

\begin{table}[H]
\centering
\caption{\textbf{Permeability and total mineral dissolution rate.} Values are means $\pm$ sample standard deviations across three training seeds (42, 1234 and 2026) at each time. $K$ is in $10^{-12}\,\mathrm{m^2}$ and $R$ in $10^{-9}\,\mathrm{mol\,s^{-1}}$. Signed errors are calculated for each seed as $100(X/X_{\mathrm{ref}}-1)$ before summarising across seeds; they retain the direction of the deviation from the reference. The standard deviations describe variation between training runs on the same sample. All simulations use the native voxel grid.}
\label{supp-tab:cfd_model_values}
\begin{tabular}{llcccc}
\toprule
Time (min) & Model & $K$ & $R$ & $\delta K$ (\%) & $\delta R$ (\%) \\
\midrule
0 & 3D & $7.765\pm0.663$ & $3.193\pm0.113$ & $+12.99\pm9.64$ & $-4.15\pm3.39$ \\
 & 3D + input corruption & $7.819\pm1.128$ & $3.270\pm0.067$ & $+13.77\pm16.42$ & $-1.82\pm2.02$ \\
 & 4D & $6.723\pm0.354$ & $3.180\pm0.041$ & $-2.18\pm5.15$ & $-4.53\pm1.24$ \\
\midrule
187 & 3D & $30.612\pm1.020$ & $3.769\pm0.023$ & $-6.71\pm3.11$ & $-3.13\pm0.59$ \\
 & 3D + input corruption & $30.239\pm1.890$ & $3.759\pm0.036$ & $-7.85\pm5.76$ & $-3.39\pm0.92$ \\
 & 4D & $32.802\pm0.223$ & $3.838\pm0.025$ & $-0.03\pm0.68$ & $-1.34\pm0.65$ \\
\midrule
646 & 3D & $46.997\pm2.032$ & $3.799\pm0.023$ & $-17.96\pm3.55$ & $-3.03\pm0.58$ \\
 & 3D + input corruption & $45.609\pm2.772$ & $3.778\pm0.036$ & $-20.38\pm4.84$ & $-3.57\pm0.91$ \\
 & 4D & $58.482\pm1.097$ & $3.853\pm0.004$ & $+2.09\pm1.92$ & $-1.67\pm0.09$ \\
\bottomrule
\end{tabular}
\end{table}

\begin{table}[H]
\centering
\caption{\textbf{Permeability relative to 0 min.} The ratio $K(t)/K(0)$ gives permeability at time $t$ relative to its initial value. Ratios are calculated within each seed before taking the mean and sample standard deviation across three training seeds. The final column is the mean and sample standard deviation of each seed\textquotesingle s absolute relative deviation from the reference ratio.}
\label{supp-tab:cfd_increase}
\begin{tabular}{llcc}
\toprule
Interval (min) & Segmentation & $K(t)/K(0)$ & Absolute error (\%) \\
\midrule
0--187 & Reference & 4.77 & --- \\
 & 3D & $3.95\pm0.22$ & $17.17\pm4.61$ \\
 & 3D + input corruption & $3.90\pm0.34$ & $18.34\pm7.08$ \\
 & 4D & $4.89\pm0.25$ & $4.78\pm1.03$ \\
\midrule
0--646 & Reference & 8.34 & --- \\
 & 3D & $6.07\pm0.31$ & $27.19\pm3.77$ \\
 & 3D + input corruption & $5.88\pm0.56$ & $29.40\pm6.74$ \\
 & 4D & $8.72\pm0.56$ & $6.73\pm3.07$ \\
\bottomrule
\end{tabular}
\end{table}
\begin{table}[H]\centering

\caption{\textbf{Flow-solver settings.} These settings apply to the fields used for visualisation. Pressure and velocity targets are the outer-iteration initial residual thresholds; velocity relaxation is dimensionless. The velocity target applies to each component. Reference denotes the reference segmentation, and 3D and 4D denote the matched models at training seed 42.}
\label{supp-tab:flow_visualisation_settings}
\begin{tabular}{lccc}
\toprule
Field & Pressure target & Velocity target & Velocity relaxation\\
\midrule
0\,min, reference & \(10^{-7}\) & \(10^{-7}\) & 0.99\\
0\,min, 3D & \(10^{-6}\) & \(10^{-4}\) & 0.90\\
0\,min, 4D & \(5\times10^{-7}\) & \(2\times10^{-5}\) & 0.99\\
\shortstack[l]{187 and 646\,min,\\all groups} & \(3\times10^{-6}\) & \(10^{-4}\) & 0.90\\
\bottomrule
\end{tabular}
\end{table}

\begin{table}[H]
\centering
\caption{\textbf{Variation across repeated random perturbations of pore boundaries.} Each range gives the minimum and maximum relative error across $n$ random draws at the same time step, using the reference calculation at that time. Main-text Fig.~\ref{fig:flow_simulation}a,b shows the corresponding means. A zero pore-volume change exchanges 3\% of the original pore-voxel count in each direction. Only settings with repeated draws are listed; all other settings used one draw.}
\label{supp-tab:random_boundary_ranges}
\begin{tabular}{ccccc}
\toprule
\shortstack{Time\\(min)} & \shortstack{Pore-volume\\change (\%)} & $n$ & \shortstack{Permeability\\error range (\%)} & \shortstack{Total mineral dissolution\\rate error range (\%)} \\
\midrule
0 & $-3$ & 3 & $[-22.650, -22.420]$ & $[12.132, 12.362]$ \\
0 & $0$ & 2 & $[-21.328, -21.039]$ & $[26.920, 26.955]$ \\
0 & $+3$ & 3 & $[4.307, 4.342]$ & $[14.787, 14.874]$ \\
187 & $-3$ & 3 & $[-18.836, -18.790]$ & $[15.992, 16.016]$ \\
187 & $0$ & 2 & $[-18.084, -17.852]$ & $[33.901, 33.904]$ \\
187 & $+3$ & 3 & $[3.800, 3.834]$ & $[17.340, 17.374]$ \\
\bottomrule
\end{tabular}
\end{table}

\Needspace{8\baselineskip}
\section{Supplementary Notes}\label{supp-sec:si_notes}
\suppnote{Dissolution segmentation accuracy and the contribution of other images}
\label{supp-sec:supp_temporal_ablation_results}
We assessed segmentation accuracy during dissolution and examined the contribution of other images in the time sequence (Supplementary Table~\ref{supp-tab:dissolution_temporal_ablation}). In the tables, 3D denotes models that segment each image independently, and 4D denotes PoreTimeNet. The dissolution sequence is described by Ma et al.~\cite{si-ma2026time}. All Seq1 dissolution results use the complete \(500^3\)-voxel evaluation region and ten time steps with reference segmentations. Each metric was averaged over time steps within each training run, then summarised across training seeds 42, 1234 and 2026. Values are means and sample standard deviations; the seeds describe training variability within one sample. Spatially and temporally correlated observations are not treated as independent samples. Fluid Dice and BF1 scores average oil and brine equally. Porosity and saturation errors are absolute differences in percentage points (pp). The additional 3D model was trained with input corruption enabled; the current-time-step-only 4D model retains temporal layers but cannot use information from other images in the input window during training or inference.

At 300\,min, the mean absolute distance from the predicted pore--rock surface to the reference surface over the complete $500^3$ evaluation region was $1.34\pm0.04$ voxels for 3D and $0.38\pm0.02$ voxels for 4D (mean and sample standard deviation across three training seeds). From 0 to 646\,min, the inferred pore-volume ratio was approximately 1.57 for the matched 3D models, compared with 1.71 for the reference segmentation and 1.71--1.76 across the 4D training seeds (mean 1.74). The current-time-step-only 4D model had rock BF1 scores 0.001--0.010 below the matched 3D model in the seed-wise comparisons.

\suppnote{Network features at phase boundaries}
\label{supp-sec:decoder_features}
We compared how clearly the network features distinguish phases at their boundaries (Supplementary Fig.~\ref{supp-fig:decoder_features}). We extracted L2-normalised, 48-channel decoder features immediately before the final classification layer at the central time-step output of each five-image window. At each time step with a reference segmentation, we sampled 300 voxels per phase at each interface from distinct $4^3$-voxel grid cells, requiring the reference segmentation to contain exactly the two specified non-background phases within the centre voxel and its 26-neighbourhood. Two-dimensional t-SNE~\cite{si-van2008visualizing} was fitted separately for each model, seed and interface using Euclidean distance, PCA initialisation, perplexity 50, an automatic learning rate and 2,000 iterations (random state 42). Phase purity was the fraction of 15 cosine-nearest neighbours in the original feature space with the same reference phase, restricting neighbours to the same interface type at other time steps and weighting the two phases equally.

\suppnote{Pore-to-rock changes in predictions and segmentation accuracy}
\label{supp-sec:d2_consistency}
We assessed whether reducing pore-to-rock changes in predictions also preserves segmentation accuracy (Supplementary Table~\ref{supp-tab:d2_consistency}). The diagnostic counts voxels whose predictions change from pore to rock between consecutive sampled time steps, divided by the number of voxels in the complete \(500^3\) evaluation region and averaged over nine intervals. It is not a segmentation-accuracy objective. A rule that permanently retains every voxel once predicted as pore would eliminate reverse transitions while retaining early false-pore predictions. Three-time-step voting provides a measured comparison: it reduces reverse transitions but also reduces the fluid Dice score. These voting values are not measurements of the irreversible rule.

\suppnote{Segmentation accuracy and consistency in Ketton limestone}
\label{supp-sec:ketton_matched}
We compared segmentation accuracy and changes in predicted rock across the Ketton drainage and imbibition sequences (Supplementary Tables~\ref{supp-tab:ketton_drainage}--\ref{supp-tab:ketton_repeated}). The drainage and imbibition images are from Singh et al.~\cite{si-singh2018time}. For both drainage and imbibition, input corruption was disabled for the 3D model and enabled for the 4D model. Additional 3D models used input corruption to assess its effect (3D + input corruption). Tables use the central time-step output; temporal-window aggregation is a separate setting. Drainage scoring covers all 125,000,000 voxels. Imbibition scoring retains 79,881,984 voxels after excluding the training cube and predictions whose inference windows touch it. Both regions include the validation region used to select these models' checkpoints. 

\suppnote{Effect of the number of labelled training time steps on segmentation}
\label{supp-sec:lux_budget}
We compared brine segmentation accuracy and phase-volume errors for Luxembourg sandstone models trained with one, three or five labelled time steps (Supplementary Tables~\ref{supp-tab:lux_budget} and~\ref{supp-tab:lux_budget_volume}). All results here use models trained from random initialisation and the combined validation and test region. N1, N3 and N5 denote one, three and five labelled training time steps; separate validation annotations were additionally used for checkpoint selection. The evaluated time steps were at 0, 5, 11, 16, 21, 27, 32, 37, 42, 47, 53, 58, 63, 69 and 74\,min, measured from the start of continuous imaging as defined in the main-text Methods. These comparisons exclude 37\,min, which belongs to every set of labelled training time steps. Brine Dice scores and volume errors were calculated using the remaining 14 images. Each metric is averaged over time for each seed and then over seeds.

\suppnote{Effect of the time interval between input images on segmentation accuracy}
\label{supp-sec:lux_input_spacing}
We tested how changing the time interval between input images affects the accuracy of the same trained 4D models, without further training (Supplementary Table~\ref{supp-tab:lux_input_spacing}). The input images were separated by one, three or five time-step intervals. These correspond to 60, 180 and 300\,s, all present during training. Predictions were compared for time steps at 11, 16, 21, 27, 32, 37, 42, 47, 53, 58 and 63\,min. These times limited each input window to at most three positions outside the image sequence, which were filled by repeating the nearest endpoint image. These common times include 37\,min and differ from those used to compare models trained with one, three or five labelled time steps; scores should therefore be compared within the same table. The 3D predictions were reused because their output does not depend on the time interval between input images.

\suppnote{Measuring oil filling between time steps}
\label{supp-sec:lux_observed_filling}
We assessed how well 3D and 4D segmentations recover oil filling observed between time steps (Supplementary Table~\ref{supp-tab:lux_filling_volume}).
The common reference pore mask contains 8,442,760 voxels in the combined validation and test region. Within this fixed mask, filling consists of voxels classified as brine at the earlier time step and oil at the later time step. Each reference or model mask is filtered independently to retain 26-connected components of at least 24 voxels; components touching a region boundary are included. The intervals between the selected time steps span five or six minutes each. Summing the filling volumes across these intervals gives the total observed brine-to-oil transition volume.

The largest retained reference filling region occurred between 32 and 37\,min and contained 666,548 voxels (0.348\,mm$^3$). It was selected by reference volume across all intervals and may span several pores. Recovery was calculated as the volume shared by this reference region and each model's retained filling mask, divided by the reference-region volume. Its intersections with the predicted filling masks recovered $62.4\pm0.7\%$ of the region for 3D and $86.6\pm0.5\%$ for 4D (three seeds). Main-text Fig.~\ref{fig:pore_dynamics}e renders seed 42 while the printed percentages report the means across three training seeds.

The local filling example in main-text Fig.~\ref{fig:pore_dynamics}d was selected from fixed reference pores with at least 85\% brine before filling, at least 60\% oil afterward and an oil increase of at least 50 percentage points. All three 3D models classified more than 30\% of the earlier pore as rock; all three 4D models classified less than 20\% as rock and more than 50\% as oil afterward. The displayed example had the largest reference oil-volume increase among 473 eligible pores. At 35\,min, the greyscale value is intermediate because filling occurs during the scan; the reference segmentation identifies the pore as brine. 

\suppnote{Predicted phases in pores during oil filling}
\label{supp-sec:lux_midscan_filling}
We examined whether changes in greyscale values during filling were associated with a pore being predicted as rock before oil was identified in the reference sequence (Supplementary Fig.~\ref{supp-fig:lux_midscan_filling} and Table~\ref{supp-tab:lux_midscan_rock}). This descriptive analysis used existing predictions and reference segmentations in the combined validation and test region. Fixed reference pores had volumes of at least 100 voxels, with a voxel side length of 8.051\,$\mu$m. For an event from $t-1$ to $t$, the reference oil fraction at $t$ was at least 0.60, the brine fraction at $t-1$ was at least 0.85, and the oil fraction increased by at least 0.50. The earlier time was at least 5\,min and the later time at most 69\,min. Each pore--time-step pair formed one record, without any model-based selection condition. The 1,603 records corresponded to 1,603 different pores with a combined volume of 1.10\,mm$^3$. This volume sums the complete selected pores, whereas Supplementary Table~\ref{supp-tab:lux_filling_volume} sums connected brine-to-oil transition voxels over five- or six-minute intervals.

For each pore, $G(s)$ was the median source greyscale value over all its voxels at time step $s$. We defined the relative greyscale position in the preceding image as
\[
x=\frac{G(t-1)-G(t)}{G(t-2)-G(t)}.
\]
Values near 1 resemble the earlier brine-filled state, and values near 0 resemble the later oil-filled state. For each image, global rock, brine and oil greyscale values were estimated using a fixed random seed and samples of up to 2,000,000 valid voxels per phase. Rock samples used valid non-pore voxels; fluid samples used pore voxels assigned to the relevant phase in the reference segmentation sequence. The rock level on the same scale was $x_{\mathrm{rock}}=[G_{\mathrm{rock}}(t-1)-G(t)]/[G(t-2)-G(t)]$, with a median of approximately 0.83 across events. We used $x<0.8$ as the operational criterion for filling during the scan; 922 pores met this criterion. These pores contained 1,144,041 of the 2,103,339 voxels in all 1,603 selected pores (54.4\%). The interval $0.6\le x<0.8$ contains 496 pores near the rock greyscale value. These intervals use a common threshold, rather than a separate threshold at each pore's $x_{\mathrm{rock}}$.

For each model, we measured the fraction of each pore predicted as rock at $t-1$. Subset summaries were weighted by pore volume, then averaged over training seeds 42, 1234 and 2026 for the 3D and 4D models. Supplementary Fig.~\ref{supp-fig:lux_midscan_filling} groups $x$ into the intervals below 0.2, 0.2--0.4, 0.4--0.6, 0.6--0.8 and at least 0.8. Temporal majority voting used the saved 3D predictions from seed 42. At each voxel, the most frequent valid predicted phase was selected over $[t-1-2,t-1+2]$, including $t-1$; background or invalid predictions (stored value 0) were excluded from voting, and ties used the prediction for the central image. The resulting volume-weighted rock fraction in the $x<0.8$ subset was 0.908, compared with 0.909 before voting for the same seed.

We also recorded the first time step within $[t-3,t+3]$ at which the predicted oil fraction reached 0.5. For seed 42, this time matched $t$ in 1,550 of 1,589 detected events for 3D and 1,579 of 1,593 for 4D; 14 and 10 events, respectively, did not reach the threshold within the window. With 3D temporal voting, 1,360 of 1,555 detected events matched $t$, 194 were later, one was earlier, and 48 did not reach the threshold. Most 3D and 4D predictions without temporal voting therefore identified oil at the reference event time, despite their different predicted phases in the pores in the preceding image. 

\suppnote{Segmentation of other dissolution experiments without retraining}
\label{supp-sec:transfer_matched}
We evaluated models trained on one dissolution sequence on two further sequences without retraining (main-text Table~\ref{tab:dissolution_transfer}).
The target sequences are described by Ma et al.~\cite{si-ma2026multiphase}. The three paired 3D and 4D models trained on the water-wet dissolution sequence (Seq1; Supplementary Note~\ref{supp-sec:supp_temporal_ablation_results}) were applied without target adaptation or target checkpoint selection to mixed-wet Seq2 and water-wet Seq3 from a second sample. Each target sequence had reference segmentations at two time steps. Seq2 contained 18 images; Seq3 contained seven images spanning a baseline and six injection time steps at 30-minute intervals. Predictions retained the source-model preprocessing and used five-image temporal windows with \(128^3\)-voxel spatial inputs and 50\% spatial overlap. Unlike the Seq1 evaluation using the central time-step output, overlapping temporal-window predictions for each time step with a reference segmentation were combined using Gaussian weights (width five time-step indices, minimum weight 0.05). Voxels outside non-zero greyscale support were masked. This comparison is restricted to overlap and boundary metrics. Metrics were calculated within valid sample support, averaged over the two time steps with reference segmentations and then summarised across seeds.

\suppnote{Comparison with nnU-Net}

\label{supp-sec:nnunet_ct}

We compared the 3D SwinUNETR model and PoreTimeNet with nnU-Net trained using its own configuration and training procedure. We trained nnU-Net v2.8.0~\cite{si-isensee2021nnu} with its 3D full-resolution configuration on the same dissolution training region and ten labelled time steps. The spatial split used the same validation region as the matched SwinUNETR comparison. The network was a six-stage convolutional U-Net with $160\times160\times96$-voxel patches. The global batch size was 16 over two GPUs. Training used 1,000 epochs of 250 iterations, Dice and cross-entropy losses with deep supervision, and Nesterov SGD with initial learning rate 0.01, momentum 0.99 and weight decay $3\times10^{-5}$, with polynomial learning-rate decay. The SwinUNETR comparisons isolate temporal processing within their shared architecture and training protocol.

Rock, oil and brine were encoded as foreground classes, with a separate background class outside the sample. Within the sample, each voxel was assigned to the most probable of the three phases during inference. Inference used sliding windows with a step size of 0.5 and Gaussian weighting, without test-time mirroring, and the same $628^3$ input context and central $500^3$ evaluation region as the dissolution comparison. Scores use the same phase mapping, valid support and metric functions as the matched models.

Supplementary Table~\ref{supp-tab:nnunet_ct_summary} reports the comparison with the models developed in this paper.

\section{Supplementary Methods}\label{supp-sec:si_methods}
\subsection{Architecture}\label{supp-sec:supp_architecture}
Supplementary Fig.~\ref{supp-fig:supp_architecture} details the architecture. The spatial encoder is shared across images. Temporal attention exchanges features at corresponding spatial locations, while decoder fusion operates at the three coarsest stages. The current-time-step-only 4D model limits the attention matrix to its diagonal; the architecture and nominal parameter count remain unchanged. Main-text equations define the attention, fusion and loss functions.

\subsection{Segmentation evaluation and shared training settings}
\label{supp-sec:supp_spatial_evaluation}
The Dice and boundary F1 scores follow main-text Eqs.~\eqref{eq:evaluation_dice}--\eqref{eq:boundary_f1}. Boundaries use six-neighbour binary erosion and an inclusive Euclidean matching tolerance of $d\le1$ voxel (the same voxel or a face neighbour). For both dissolution and Ketton displacement, phases are restricted to the evaluation region and valid sample support before boundary extraction, with zero padding outside them. For these datasets, Dice and BF1 scores are zero when only one of the two sets is empty; when both are empty, the score is undefined and omitted from the mean. Because erosion treats voxels outside the evaluation region, valid support or array as background, voxels on the faces of the evaluation regions, on the faces of the excluded imbibition cube \([144,500)^3\) and on the edge of the Seq2 and Seq3 sample support count as boundary voxels. The same boundary definition applies to both architectures, without an edge correction.

Dissolution inference used a \(628^3\)-voxel greyscale input extending 64 voxels beyond each face of the \(500^3\)-voxel evaluation region. The context includes greyscale values from the validation region but no validation labels or training region. Ketton predictions were assembled on a \(500^3\)-voxel grid using \(96^3\) spatial windows with 50\% overlap. Imbibition excludes \([144,500)^3\), including the \([200,500)^3\) training cube and touching inference windows. Luxembourg sandstone predictions use \(96^3\) windows with 50\% spatial overlap and Gaussian blending over the $725\times728\times256$-voxel evaluation region. Its training, validation and test regions are adjacent, with no gap.

For Ketton drainage and imbibition, three-phase training annotations occupied a \(300^3\)-voxel region within each image. The labelled training time steps were \(\{1,50,100,200,350,460\}\) for drainage and \(\{1,100,170,200,230,380\}\) for imbibition.

The main-text Methods describe the shared training procedure, and Supplementary Table~\ref{supp-tab:dataset_settings} summarises the settings by dataset.

Shared spatial augmentation used axis flips with probability 0.5, affine transforms with probability 0.3 (rotation range 0.2618 rad and scale range 0.15), and Gaussian smoothing with probability 0.1 (standard deviation 0.5--1.0 voxels). Temporal intensity drift had probability 0.5, maximum scale/shift 0.15/0.10 and per-time-step scale/shift standard deviations 0.03/0.02. Temporal reversal had probability 0.5 for the displacement datasets and was disabled for dissolution. When input corruption was enabled, a selected training image received no change, Gaussian noise, cuboid masking or intensity jitter with probabilities 0.2, 0.2, 0.4 and 0.2. Noise standard deviation was 0.15 in scaled greyscale values; cuboid masks had side fractions from 0.2 to 0.4 and were filled using a blurred image at the current time step (five-voxel kernel) and noise 0.02. Intensity jitter used scale 0.75--1.25 and shift \(-0.15\) to \(0.15\), followed by clipping. Only the masking branch of input corruption could also perturb other images in the input window. It selected zero to two images within two positions of the selected image in the input window and applied weaker perturbations to the same cuboid. Disabling input corruption also disabled these perturbations. These augmentations were also used during training on the Luxembourg sandstone dataset; input corruption was enabled for its 4D models and disabled for its 3D models.

\subsection{Luxembourg sandstone training and inference}
\label{supp-sec:lux_training_settings}
The Luxembourg sandstone analysis compared 3D and 4D models trained with one, three or five labelled time steps. Each combination was trained from random initialisation with seeds 42, 1234 and 2026, giving 18 models. The labelled time steps were at 37\,min for N1; 18, 37 and 56\,min for N3; and 9, 18, 37, 56 and 65\,min for N5. Validation used time steps at 0, 16, 27, 32, 37, 47, 58, 69 and 74\,min in the separate validation region.

Input greyscale values were clipped to the range \([19{,}616,47{,}547]\) and mapped linearly to \([0,1]\). Each training window contained \(96^3\)-voxel patches from the same spatial region in 11 images. Time intervals of 60, 180 and 300\,s between these images were sampled equally. Rock, oil and brine crop centres were sampled with equal weights. Predictions at every time step with training labels in the window contributed to the segmentation loss.

The training used 115 batches per epoch, mixed precision and a gradient-norm limit of 1. Input corruption was enabled for 4D and disabled for 3D. The weaker perturbations applied to other images in the input window during masking used a blur kernel of three voxels, noise standard deviation 0.03, intensity scale 0.95--1.05 and shift \(-0.03\)--0.03.

We used sliding-window inference with \(96^3\)-voxel patches, 50\% spatial overlap and Gaussian weighting to combine predictions from overlapping patches. Input images for the 4D comparisons were separated by 60\,s, except when testing the input interval (Supplementary Note~\ref{supp-sec:lux_input_spacing}). We retained the prediction for the central time step of each input window. Where the window extended beyond the sequence, the nearest endpoint image was repeated.

\subsection{Solver convergence and tolerance checks}
\label{supp-sec:cfd_numerics}
Supplementary Tables~\ref{supp-tab:cfd_model_values} and~\ref{supp-tab:cfd_increase} summarise permeability, total mineral dissolution rate and permeability increases across training seeds. The 30 flow calculations for the model comparison used SIMPLEC with velocity relaxation 0.90. They included the reference geometries and predictions from the 3D model trained without input corruption, the 3D model trained with input corruption, and the 4D model. Convergence required final outer-iteration initial residuals of \(p\leq5\times10^{-7}\) and each velocity component \(\leq2\times10^{-5}\). Absolute and relative linear-solver tolerances were \(10^{-12}\) and 0.01. Both inlet and outlet flow rates had to agree with the prescribed flow rate to within \(10^{-6}\), relative to that rate. All 30 cases met these criteria. One 187-min 4D geometry required continuation of its existing solution with unchanged numerical settings; no flow calculation was accepted solely because it reached an iteration limit.

The transport solver used an unnormalised concentration-residual target of \(QC_{\mathrm{in}}10^{-6}\), absolute linear tolerance \(QC_{\mathrm{in}}10^{-10}\), and relative tolerance 0.01; these expressions use the solver's concentration units of \(\mathrm{kmol\,m^{-3}}\), with \(C_{\mathrm{in}}=0.025\,\mathrm{kmol\,m^{-3}}\). An alternative acceptance criterion required at least 60 iterations and a 50-iteration window with residuals above the target but with the total mineral dissolution rate, \(R\), remaining stable. The ratio of maximum to minimum residual minus one could not exceed 0.05, or 0.50 if every residual in the window was within ten times the target. The corresponding ratio of the maximum to minimum \(R\), minus one, had to remain below \(10^{-5}\), using at least three samples of \(R\). Of the 30 cases, 26 reached the residual target and four met the alternative criterion; all four were 0-min geometries. For one of these cases, the final iteration window was assessed against this criterion after the prescribed iteration limit was reached.

For both acceptance routes, each case was advanced for three additional iterations and had to pass the same checks. Relative water imbalance was limited to \(10^{-6}\), solute imbalance to \(10^{-4}\), and disagreement between \(R\) and rock-wall diffusive flux to \(10^{-6}\). Concentrations had to remain finite and within \([-0.001,1.001]C_{\mathrm{in}}\). Over the additional iterations, mean and maximum absolute concentration changes were limited to \(10^{-6}C_{\mathrm{in}}\) and \(10^{-4}C_{\mathrm{in}}\), respectively, and relative change in \(R\) to \(10^{-5}\). Wall flow and diffusive flux through non-reactive sidewalls were also checked against zero.

A tighter-tolerance check considered six geometries: the reference and 3D seed 42 at 0\,min, the reference and 4D seed 2026 at 187\,min, and the reference and 3D seed 42 at 646\,min. Pressure and velocity targets were \(10^{-7}\) and \(5\times10^{-6}\); transport relative tolerance was 0.001; the minimum iteration count and window used to assess stability of \(R\) were 120 and 100; and thresholds for relative changes in \(R\) were \(10^{-6}\). Both 0-min cases and the 187-min reference met these tighter criteria, with permeability and \(R\) differing from the corresponding calculations in the model comparison by at most 0.005\%. The other three did not meet the tighter criteria and were not counted as successful checks.

The 30 calculations for the model comparison, the nnU-Net calculations, the random boundary perturbations and the fields used for visualisation all met common flow criteria: final outer-iteration initial residuals of \(p\leq3\times10^{-6}\) and \(\leq10^{-4}\) for each velocity component, and relative inlet/outlet flow-rate error \(\leq10^{-5}\). All passed the concentration, solute-balance, wall-flux and additional-iteration stability checks described above; relative water imbalance was at most \(10^{-5}\). These checks were applied to the recorded final solutions; detailed settings for the perturbation and visualisation calculations are given below.

\subsection{Random perturbations to pore boundaries}
\label{supp-sec:random_pore_boundary_changes}
Random boundary changes were made separately to the reference segmentation at 0, 187 and 646\,min. Candidate voxels belonged to the one-voxel layer on either side of the rock--pore interface, using face neighbours only. To increase pore volume, rock voxels sharing a face with pore space were selected uniformly without replacement and reclassified as brine. To decrease pore volume, pore voxels sharing a face with rock were selected in the same way and reclassified as rock. For a requested fraction \(f\), the number selected was \(\operatorname{round}(fN_{\mathrm{pore,ref}})\), where \(N_{\mathrm{pore,ref}}\) counts all oil and brine voxels in the reference region before extraction of the spanning flow domain. Fractions were 1, 3 and 5\% in each direction. Exterior voxels were unchanged.

For random perturbations that preserve total pore volume, \(\operatorname{round}(0.03N_{\mathrm{pore,ref}})\) voxels were changed in each direction, with both sets drawn from the original boundary layers. Thus the total pore volume was preserved exactly, although its geometry and the volume of the spanning flow domain could change. Each altered segmentation was processed by the same connected-domain extraction and native-grid meshing procedure used for model predictions.

At 0 and 187\,min, the 3\% changes had three independent random draws per direction, the net-zero change had two, and the 1\% and 5\% changes had one per direction, giving 12 cases at each time step. At 646\,min there was one draw for each of these seven conditions, giving 31 cases overall. Repeated draws were averaged for the plotted points; they are separate boundary realisations, not additional model-training seeds. The porosity error in these plots is the relative pore-volume difference, \(100(N_{\mathrm{pore}}/N_{\mathrm{pore,ref}}-1)\), rather than a difference in porosity percentage points. The ranges across repeated draws are reported in Supplementary Table~\ref{supp-tab:random_boundary_ranges}.

These 31 cases used flow targets \(p\leq3\times10^{-6}\), each velocity component \(\leq10^{-4}\), velocity relaxation 0.90, and relative flow-rate error \(\leq10^{-5}\). All passed flow checks and reached their transport residual target. Physical parameters and boundary conditions were unchanged. Their plotted errors use the same reference values as the model comparison. Replacing the reference values calculated with the earlier tolerances by these reference values changed the errors for the randomly perturbed geometries by at most 0.101 percentage points for permeability and 0.015 percentage points for \(R\).

Across the three time steps, net-zero boundary changes reduced permeability by 16.9--21.2\% and increased the total mineral dissolution rate by 26.9--36.4\%. Repeated random draws were averaged at each time step before these ranges were reported.

\subsection{Flow and concentration fields used for visualisation}
\label{supp-sec:flow_visualisation_settings}
The field visualisations show one model per architecture (seed 42) and the corresponding reference geometry at each time step. They use the same segmentation, voxel grid, physical parameters and boundary conditions as the corresponding calculations in the model comparison, but were retained or recalculated with the flow tolerances in Supplementary Table~\ref{supp-tab:flow_visualisation_settings}. Numerical values in the comparisons of permeability and \(R\) use the 30 calculations described above.

\renewcommand{\refname}{Supplementary References}\phantomsection\label{supp-sec:si_references}

\clearpage\endgroup
\end{document}